\documentclass[aps,12pt,eqsecnum]{revtex4}
\usepackage{amsmath}
\usepackage{graphicx}
\newcommand{\beq}{\begin{equation}}
\newcommand{\eeq}{\end{equation}}
\newcommand{\bes}{\begin{subequations}}
\newcommand{\ees}{\end{subequations}}
\newcommand{\bea}{\begin{eqnarray}}
\newcommand{\eea}{\end{eqnarray}}
\newcommand{\kk} {~}
\newcommand{\kko} {~~~~~~~~~~~~~~~~~~~~}

\newcommand{\alp}{\mbox{\boldmath$\alpha$}}

\begin{document}
\title{Coulomb Excitation of Multi-Phonon Levels of The Giant Dipole 
Resonance} 
\author{B.F. Bayman\\
{\it School of Physics and Astronomy, University of Minnesota,}\\
{\it Minneapolis, MN, U.S.A.}\\
and\\
F. Zardi\\
{\it Istituto Nazionale di Fisica Nucleare and Dipartimento di Fisica,} \\
{\it Padova,Italy.}}
\begin{abstract}
A closed expression is obtained for the cross-section for Coulomb excitation of levels of the giant dipole resonance of given angular momentum and phonon number. Applications are made to the Goldhaber-Teller and Steinwedel-Jensen descriptions of the resonance, at non-relativistic and relativistic bombarding energies.
\end{abstract}
\maketitle
\section{Introduction}
The giant dipole resonance (GDR) is one of the best-studied collective modes of nuclear excitation. It has been modeled, macroscopically, as a bulk oscillation of neutrons relative to protons \cite {GT}, or as local isovector fluctuations of neutron and proton fluids \cite {SJ}. It can also be modeled, microscopically, in terms of isovector linear combinations of particle-hole excitations of the nuclear ground state (\cite {LA,CA,BP,ABE}). It has been studied experimentally using reactions induced by gamma rays, light ions, and by the sharp pulses of electromagnetic radiation associated with projectile nuclei moving at relativistic speeds, {\it i.e.} relativistic Coulomb excitation. A broad survey of the GDR and other resonances, and different different excitation methods, is given in ref.\cite{CF}. In this paper, we will consider relativistic Coulomb excitation of GDR states which are described by macroscopic models.

In many situations, the de Broglie wave length for the  relative motion of the projectile and target in a Coulomb excitation experiment is small compared to the linear dimensions that characterize the system. Then a semi-classical approach can be used, in which this relative motion is described in terms of a classical orbit, whereas the internal changes of the projectile and target are described using quantum mechanics. We are interested in a situation in which the projectile remains unexcited, but the target is excited to the states associated with the GDR. 

A useful first approximation to an oscillatory situation, such as the GDR, is to assume that the restoring forces are proportional to the displacement from equilibrium. With this approximation, the GDR is dynamically equivalent to an isotropic 3-dimensional harmonic oscillator. This assumption leads to the familiar harmonic oscillator spectrum, in which eigenstates are characterized by phonon number, total angular momentum, and angular-momentum-z-component. If the oscillator picture were exact, all the $(N+1)(N+2)/2$ eigenstates with $N$ phonons would be degenerate in energy. Deviations from the oscillator picture would lift this degeneracy, but if the deviations were spherically symmetric we would still have the $2\ell+1$-fold degeneracy of the angular-momentum eigenstates with $\ell=N, N-2, N-4,\ldots,0$ or 1.

The coupling between the electromagnetic pulse due to the projectile and the internal oscillating degrees of freedom associated with the GDR of the target can usefully be approximated by an expression that is linear in these oscillating degrees of freedom. If this approximation is made, an exact solution can be found for the Schr\"{o}dinger equation that describes the time evolution of the target \cite{BE,ME}. Formulae have been published in the literature for the total excitation probability of all GDR states of given phonon number, when the relative motion of the target and projectile is along a specified orbit. In this paper, we decompose this total excitation probability into the contributions of phonon states of given total angular momentum. For example, we show how to find the excitation probabilities of four-phonon states of angular momentum 0, or 2, or 4, whereas the previously published formula yielded only the total four-phonon excitation probability.

The model studied in this paper is highly simplified, since it uses a pure oscillator description of the GDR, and a linear approximation of the coupling between the projectile motion and the GDR degrees of freedom. More realistic calculations have been done, for example in refs. \cite{LA,CA,BP,ABE,CF},
 which have proven useful in the analysis of a wide variety of Coulomb excitation experiments. We believe that our simplified model may, nevertheless, be useful in indicating some general trends, especially with respect to the variation of excitation cross-section with bombarding energy, and the way this depends upon the angular momentum of the final state.

Unfortunately, these GDR phonon states of specified angular momentum are not clearly resolved in the excitation spectra. Indeed, superposed on the multiphonon GDR states are collective excitations of other characters, such as giant quadrupole and giant octupole excitations (see, {\it e.g.,} ref. \cite{LA}). Thus we cannot check our predictions for excitation cross-sections of GDR states of given angular momentum against any currently available data. However, it is possible that future measurements of angular distributions of the decay products of GDR states will give information about the angular momenta of these states. For example, the gamma rays emitted by the $\ell=0$ member of the two-phonon sextuplet will have a spherically symmetric angular distribution, whereas the five $\ell=2$ members will emit gamma rays with quadrupole and hexadecapole distributions. In situations such as this, it will be important to be able to predict the excitation cross-sections of $N-$phonon states of specified angular momentum.

\section{Excitation probabilities}
\subsection{The coupled Equations}
The time-dependent Schr\"{o}dinger equation for the perturbed target wave function
$\Psi(t)$ is
\beq
i\hbar \frac{\partial \Psi(t)}{\partial t}=[H_0+V(t)]\Psi(t)
\eeq
with $H_0$ the unperturbed target Hamiltonian, whose eigenfunctions are 
$\psi_\alpha$ and eigenvalues $\epsilon_\alpha$. The solution of this equation is expressed in terms of occupation amplitudes $a_\alpha(t)$ which occur in the
expansion
\beq
\Psi(t)=\sum_\alpha a_\alpha(t)e^{-\frac{i}{\hbar}\epsilon_\alpha t}\psi_\alpha.
\eeq
Substitution of (2.2) into (2.1) yields a set of coupled differential equations
\beq
i\hbar \dot{a}_\alpha(t)=\sum_\beta e^{\frac{i}{\hbar}(\epsilon_\alpha-\epsilon_\beta) t}<\psi_\alpha|V(t)|\psi_\beta>a_\beta(t).
\eeq
The intial conditions appropriate to a typical nuclear reaction are
$ a_\alpha(-\infty)=\delta_{\alpha,0}$, corresponding to the requirement
that the target be in its ground state at the start of the process. The probability that the reaction leaves the target in the final state $\psi_\alpha$
is then $| a_\alpha(\infty)|^2$.

In a Coulomb excitation reaction, the perturbation matrix elements
required in (2.3) are \cite{JA,BZ}
\beq
<\psi_\beta|V(t)|\psi_{\alpha}>~=~\int~[\varphi_{_C}^{\rm ret}({\bf r},t)\rho_{\beta \alpha}({\bf r})
~-~\frac{1}{c}{\bf A}_{_C}^{\rm ret}({\bf r},t)\cdot{\bf J}_{\beta \alpha}({\bf r}
)]d^3r~~.
\eeq
Here $\varphi_{_C}^{\rm ret}({\bf r},t)$ and ${\bf A}_{_C}^{\rm ret}({\bf r},t)$
are, respectively, the scalar and vector potentials associated with the
electromagnetic field created by the charged projectile. 
The properties of the target states $\psi_\alpha$ and $\psi_\beta$ are expressed
in (2.4) by the transition charge density $\rho_{\beta \alpha}({\bf r})$ and
current density ${\bf J}_{\beta \alpha}({\bf r})$.
\subsection{Excitation probabilities in a Cartesian basis}
In this paper we will be concerned with situations in which the target
Hamiltonian $H_0$ is that of an isotropic three-dimensional harmonic oscillator
with reduced mass $M$ and natural frequency $\omega$. The state labels used in Sec.II.A
can be taken to represent the triplet of quantum numbers $(n_x,n_y,n_z)$,
specifying the numbers of oscillator quanta in the $x,y$ and $z$ directions.
Furthermore, we will be working in the regime in which the interaction matrix elements (2.4) can be approximated by
\beq
<\psi_{n_x,n_y,n_z}|V(t)|\psi_{n'_x,n'_y,n'_z}>~=~
-\int d^3R ~\psi^*_{n_x,n_y,n_z}({\bf R})[{\bf F}(t)\cdot{\bf R}+
{\bf G}(t)\cdot{\bf P}] \psi_{n'_x,n'_y,n'_z}({\bf R})
\eeq
where ${\bf R}(=X,Y,Z)$ represents the degrees of freedom undergoing harmonic oscillations and ${\bf P}$ represents the conjugate momenta.

Since both $H_0$ and $V$ separate in Cartesian coordinates, the problem reduces to one-dimensional forced oscillations in the $X,Y,Z$ directions, and the
occupation amplitudes $a_{n_x,n_y,n_z}(t)$ factorize
\beq
a_{n_x,n_y,n_z}(t)~=~b_{n_x}(t)~b_{n_y}(t)~b_{n_z}(t).
\eeq
In particular, the $b_{n_x}(t)$ satisfy
\bea
i\hbar\dot{b}_{n_x}(t)&=&-\sum_{n'_x}e^{\frac{i}{\hbar}(n_x-n'_x)t}<n_x|F_x(t)X+G_x(t)P_x|n'_x>b_{n'_x}(t)\nonumber\\
&=&e^{i\omega t}\left(\sqrt{\frac{\hbar}{2M\omega}}F_x(t)+i\sqrt{\frac{M\hbar\omega}{2}}G_x(t)\right)\sqrt{n}~b_{n-1}(t)\nonumber\\
&+&e^{-i\omega t}\left(\sqrt{\frac{\hbar}{2M\omega}}F_x(t)-i\sqrt{\frac{M\hbar\omega}{2}}G_x(t)\right)\sqrt{n+1}~b_{n+1}(t),
\eea
which must be solved with the intial condition $b_{n_x}(-\infty)=\delta_{n_x,0}$. It is not difficult to verify that (2.7) is satisfied by
\bes
\beq
b_{n_x}(t)=\frac{[\alpha_x(t)]^{n_x}}{\sqrt{n_x!}}e^{\beta_x(t)}
\eeq
with
\beq
\alpha_x(t)=i\int_{-\infty}^t dt'\left[\frac{F_x(t')}{\sqrt{2M\hbar\omega}}+i\sqrt{\frac{M\omega}{2\hbar}}G_x(t')\right]e^{i\omega t'}
\eeq
\beq
\beta_x(t)=i\int_{-\infty}^t dt'\left[\frac{F_x(t')}{\sqrt{2M\hbar\omega}}-i\sqrt{\frac{M\omega}{2\hbar}}G_x(t')\right]\alpha_x(t')e^{-i\omega t'}
\eeq
\ees
from which it follows that $\beta_x(t)+\beta^*_x(t)=-|\alpha_x(t)|^2$. 
{\it Here and in the following, we shall use the notation $\alpha_i$, without explicit time-dependence $(t)$, to denote the quantities $\alpha_i(\infty)$; the same convention will be also used for the quantities $a_i$ and $b_i$}. Then the probability of populating the final target state 
$\psi_{n_x,n_y,n_z}$ is
\bea
{\cal P}_{n_x,n_y,n_z}&=&|a_{n_x,n_y,n_z}|^2=|b_{n_x}|^2|b_{n_y}|^2|b_{n_z}|^2\nonumber\\
&=&\frac{(|\alpha_x|^2)^{n_x}(|\alpha_y|^2)^{n_y}(|\alpha_z|^2)^{n_z}}{n_x!n_y!n_z!}
e^{-(|\alpha_x|^2+|\alpha_y|^2+|\alpha_z|^2)}
\eea
in which $\alpha_y$ and $\alpha_z$ are defined as in (2.8b), but using $F_y(t'), G_y(t')$ and $F_z(t'), G_z(t')$, instead of
$F_x(t'), G_x(t')$. Note that the``Poisson distribution"  result (2.9) involves ``on-shell" Fourier transforms of $F_x(t), G_x(t)$, as given by (2.8b).
\subsection{Excitation probabilities in a spherical basis}
The Cartesian result (2.9) is well known \cite{BE}. However specifying the eigenstates of $H_0$ in terms of $n_x,n_y,n_z$ is not as convenient as using principal and angular momentum quantum numbers $n,\ell,m$. The advantage of using $n,\ell,m$ is that a spherically-symmetric deviation from a perfect
harmonic oscillator Hamiltonian will not mix the states labeled by different $(\ell,m)$, nor will it split the $2\ell+1$ states with different $m$-values and the same $\ell$. However, the quantum numbers $n_x,n_y,n_z$ are only useful for a perfect oscillator. Since we cannot expect
perfection in a harmonic description of the GDR, but we can expect spherical symmetry, it would be advantageous to use oscillator eigenstates characterized by $n,\ell,m$ rather than $n_x,n_y,n_z$.

Using (2.2), (2.6) and (2.8) we can write the exact time-dependent target wave function as
\beq
\Psi(t)~ =~ \sum_{n_x,n_y,n_z}\frac{[\alpha_x(t)]^{n_x}[\alpha_y(t)]^{n_y}
[\alpha_z(t)]^{n_z}}{\sqrt{n_x!n_y!n_z!}}e^{\beta_x(t)+\beta_y(t)+\beta_z(t)}
e^{-i\omega(n_x+n_y+n_z+3/2)t}~|n_x,n_y,n_z>
\eeq
where $|n_x,n_y,n_z>$ represents a harmonic oscillator eigenstate with $n_x,n_y,n_z$ quanta in the $x,y,z$ directions, respectively. It is
convenient to represent the oscillator eigenstates in terms of boson creation operators $c_x^+, c_y^+, c_z^+$ acting on a normalized vacuum (ground state), $|0>$:
\beq
|n_x,n_y,n_z>~\equiv~ \frac{(c_x^+)^{n_x}(c_y^+)^{n_y}(c_z^+)^{n_z}}
{\sqrt{n_x!n_y!n_z!}}|0>
\eeq
All the properties of these oscillator eigenstates can be obtained algebraically, starting from
\bes
\beq
<0|0>=1
\eeq
\beq
[c_\mu,c^+_\nu]=\delta_{\mu,\nu},~~~~[c_\mu,c_\nu]=[c^+_\mu,c^+_\nu]=0
\eeq
\beq
c_\mu|0>=<0|c^+_\mu=0
\eeq
\ees

If we substitute $|n_x,n_y,n_z>$ from eq.(2.11) into $\Psi(t)$ given by eq.(2.10), we get 
\bea
\Psi(t)&=&e^{\beta_x(t)+\beta_y(t)+\beta_z(t)-i\frac{3}{2}\omega t} \sum_{n_x,n_y,n_z}
\frac{[e^{-i\omega t}\alpha_x(t)c^+_x]^{n_x}[e^{-i\omega t}\alpha_y(t)c^+_y]^{n_y}[e^{-i\omega t}\alpha_z(t)c^+_z]^{n_z}}
{n_x!n_y!n_z!} |0>\nonumber\\
&=&e^{\beta_x(t)+\beta_y(t)+\beta_z(t)-i\frac{3}{2}\omega t}\sum_{N=0}^\infty
\frac{[e^{-i\omega t}(\alpha_x(t)c^+_x+\alpha_y(t)c^+_y+\alpha_z(t)c^+_z)]^N}{N!} |0>\nonumber\\
&=&e^{\beta_x(t)+\beta_y(t)+\beta_z(t)-i\frac{3}{2}\omega t}\times e^{e^{-i\omega t}(\alpha_x(t)c^+_x+\alpha_y(t)c^+_y+\alpha_z(t)c^+_z)}
\eea

We now use the polynomial identity
\beq
e^{{\bf a\cdot b}}\kk=\kk 4\pi \sum_{n,\ell}\frac{[(a_x^2+a_y^2+a_z^2)(b_x^2+b_y^2+b_z^2)]^n}{(2n)!!(2n+2\ell+1)!!}
\sum_{m=-\ell}^{\ell}(-1)^m{\cal Y}_{-m}^\ell({\bf a}){\cal Y}_{m}^\ell({\bf b})
\eeq
which is proven in Appendix A. Here the solid harmonic ${\cal Y}_{m}^\ell({\bf a})\equiv a^\ell Y_{m}^\ell({\hat a})$ is a homogeneous polynomial of degree $\ell$ in $a_x,a_y,a_z$ with the same rotational transformation properties as the spherical harmonic $Y_{m}^\ell({\hat a})$. With the help of eq.(2.14), we can rewrite eq.(2.13) in the form
\bes
\bea
\Psi(t)&=&e^{\beta_x(t)+\beta_y(t)+\beta_z(t)-i\frac{3}{2}\omega t}\sum_{n,\ell}e^{-i\omega(2n+\ell)t}\nonumber\\
&\times&\left[(-1)^n\sqrt{4\pi}\frac{[\alpha^2_x(t)+\alpha^2_y(t)+\alpha^2_z(t)]^n}{\sqrt{(2n)!!(2n+2\ell+1)!!}}\right.
\left.{\cal Y}_{-m}^\ell\left(\alpha_x(t),\alpha_y(t),\alpha_z(t)\right)|n\ell m>\right]
\eea
where
\beq
|n\ell m>\kk\equiv\kk(-1)^n\sqrt{4\pi}\frac{[(c^+_x)^2+(c^+_y)^2+(c^+_z)^2]^n}{\sqrt{(2n)!!(2n+2\ell+1)!!}}{\cal Y}^\ell_m(c^+_x,c^+_y,c^+_z)|0>
\eeq
\ees

Since $|n\ell m>$ is a homogeneous polynomial of degree $2n+\ell$ in $c^+_x,c^+_y,c^+_z$ acting on the oscillator ground state, it must be an eigenstate of the harmonic oscillator Hamiltonian with $2n+\ell$ quanta. Its rotational transformation properties are the same as the spherical harmonic $Y_m^\ell$. Moreover, it is shown in Appendix B that the numerical factor in (2.15b) guarantees that $|n\ell m>$ is normalized. Therefore, eq.(2.15a) implies that
\bea
<n\ell m|\Psi(t)>&=&e^{\beta_x(t)+\beta_y(t)+\beta_z(t)-i\frac{3}{2}\omega t}\times (-1)^m\nonumber\\
&\times&(-1)^n\sqrt{4\pi}\frac{[\alpha^2_x(t)+\alpha^2_y(t)+\alpha^2_z(t)]^n}{\sqrt{(2n)!!(2n+2\ell+1)!!}}{\cal Y}_{-m}^\ell\left(\alpha_x(t),\alpha_y(t),\alpha_z(t)\right)
\eea
is the amplitude that $|n\ell m>$ is occupied at time $t$.

The quantity of interest for the interpretation of experimental data is the sum
of the excitation probabilities of all states of given $n,\ell$:
\bea
{\cal P}_{n,\ell}& \equiv& \sum_{m=-\ell}^\ell|<n\ell m|\Psi(\infty)>|^2~ =~ 
\frac{4\pi [|\alpha_x|^2+|\alpha_y|^2+|\alpha_z|^2]^{2n}}
{(2n)!!(2n+2\ell+1)!!} \nonumber\\
&\times & e^{-(| \alpha_x|^2+|\alpha_y|^2+|\alpha_z|^2)}
\sum_{m=-\ell}^\ell|{\cal Y}^\ell_{m}(\alpha_x,\alpha_y,\alpha_z)|^2
\eea
In evaluating this sum, care must be taken when complex conjugating the solid
harmonics because the arguments $\alpha_x,\alpha_y,\alpha_z$ may be complex. We have
\bea
&&\sum_{m=-\ell}^\ell{\cal Y}^{*\ell}_{m}(\alpha_x,\alpha_y,\alpha_z){\cal Y}^\ell_{m}(\alpha_x,\alpha_y,\alpha_z)
\nonumber\\
&=&(-1)^m \sum_{m=-\ell}^\ell{\cal Y}^{\ell}_{-m}(\alpha^*_x,\alpha^*_y,\alpha^*_z){\cal Y}^\ell_{m}(\alpha_x,\alpha_y,\alpha_z)
\eea
For any two vectors ${\bf r}_1$ and ${\bf r}_2$, the spherical harmonic addition theorem gives
$$
\sum_{m=-\ell}^\ell(-1)^m~{\cal Y}^\ell_{-m}(x_1,y_1,z_1){\cal Y}^\ell_m(x_2,y_2,z_2)~ =~ 
\left[(x_1^2+y_1^2+z_1^2)(x_2^2+y_2^2+z_2^2)\right]^{\ell/2} \sum_{m=-\ell}^\ell (-1)^m Y^\ell_{-m}(\hat{ r}_1)Y^\ell_m(\hat{ r}_2)
$$
\beq
=~ \left[(x_1^2+y_1^2+z_1^2)(x_2^2+y_2^2+z_2^2)\right]^{\ell/2}\frac{2\ell+1}{4\pi}\times P_\ell\left(\frac{x_1x_2+y_1y_2+z_1z_2}{\sqrt{(x_1^2+y_1^2+z_1^2)(x_2^2+y_2^2+z_2^2)}}\right)
\eeq
The argument of the Legendre polynomial in (2.19) is just $\hat {r}_1\cdot \hat {r}_2$. Equation (2.19) can be regarded as a polynomial identity in the six variables $x_1,y_1,z_1,x_2,y_2,z_2$, which we can apply to the particular choice
$$
x_1~ \equiv~ \alpha^*_x,~~~ y_1~\equiv~\alpha^*_y,~~  z_1~\equiv~ \alpha^*_z
$$
$$
x_2~ \equiv~ \alpha_x,~~ y_2~ \equiv~ \alpha_y,~~  z_2~ \equiv~ \alpha_z
$$
Then (2.18) and (2.19) yield
$$
\sum_{m=-\ell}^\ell~ |{\cal Y}^\ell_{m}(\alpha_x,\alpha_y,\alpha_z)|^2 
$$
$$
=~ |(\alpha_x)^2+(\alpha_y)^2+(\alpha_z)^2|^\ell
\times\frac{2\ell+1}{4\pi}
P_\ell\left(\frac{|\alpha_x|^2+|\alpha_y|^2+|\alpha_z|^2}{|(\alpha_x)^2+(\alpha_y)^2+(\alpha_z)^2|}\right)
$$
and eq.(2.17) gives
\begin{eqnarray}
{\cal P}_{n,\ell}&=&\frac{2\ell+1}{(2n)!!(2n+2\ell+1)!!}\left|(\alpha_x)^2+(\alpha_y)^2+(\alpha_z)^2\right|^{2n+\ell}\nonumber\\
&&e^{-(|\alpha_x|^2+|\alpha_y|^2+|\alpha_z|^2)}
\times P_\ell\left(\frac{|\alpha_x|^2+|\alpha_y|^2+|\alpha_z|^2}{|(\alpha_x)^2+(\alpha_y)^2+(\alpha_z)^2|}\right)
\end{eqnarray}
for the total excitation probability of states with given $n,\ell$

As a simple example of the application of (2.20), consider the two-phonon states 
 $(n=0,\ell=2)$ and $(n=1,\ell=0)$. Using $P_0(x)=1$, 
$P_2(x)=(3x^2-1)/2$, we find that 
\bes
\beq
{\cal P}_{1,0}=\frac{1}{6}|(\alpha_x)^2+(\alpha_y)^2+(\alpha_z)^2|^2~e^{-(|\alpha_x|^2+|\alpha_y|^2+|\alpha_x|^2)}
\eeq
\bea
{\cal P}_{0,2}&=&\frac{(|\alpha_x|^2+|\alpha_y|^2+|\alpha_z|^2)^2-\frac{1}{3}|(\alpha_x)^2+(\alpha_y)^2+(\alpha_z)^2|^2}{2}
\nonumber\\
&\times& e^{-(|\alpha_x|^2+|\alpha_y|^2+|\alpha_x|^2)}
\eea
\ees
The sum of (2.21a) and (2.21b), the total excitation probability for two-phonon states, is seen to be
$$
{\cal P}_{1,0}+{\cal P}_{0,2}~ =~ \frac{\left(|\alpha_x|^2+|\alpha_y|^2+|\alpha_z^2|\right)^2}{2}\times e^{-(|\alpha_x|^2+|\alpha_y|^2+|\alpha_z|^2)}
$$
which is the same as the sum of ${\cal P}_{n_x,n_y,n_z}$ of (2.9) for the six $n_x,n_y,n_z$ combinations for which $n_x+n_y+n_z=2$

In the application of (2.20) to Coulomb excitation of the GDR, we will find that $\alpha_x=0$, $\alpha_y$ is  real and
$\alpha_z$ is pure imaginary. This is because the influence of the projectile in the ${\bf z}$-direction is opposite at $\pm t$,
whereas its influence in the ${\bf y}$-direction is the same at $\pm t$. This different time behavior leads to different behavior under complex conjugation of the Fourier transforms that determine $\alpha_y,\alpha_z$. The significance of the opposite signs of 
$(\alpha_y)^2$ and $(\alpha_z)^2$ is evident in (2.20). 

In general, $\alpha_x,\alpha_y$ and $\alpha_z$ will be functions of the impact parameter, $b$, which characterizes the projectile orbit. Therefore ${\cal P}_{n,\ell}$ will also be a function of $b$. The excitation cross-section involves an integral over impact parameter,
\beq
\sigma_{n,\ell}=2\pi\int_{{\rm b_{min}}}^\infty {\cal P}_{n,\ell}(b)~b~db.
\eeq 
The lower limit, $b_{{\rm min}}$, is of the order of the sum of the radii of the projectile and target nuclei \cite{ZB}. Because the electromagnetic pulse due to the projectile becomes more adiabatic as $b$ increases, ${\cal P}_{n,\ell}(b)$ decreases strongly for large $b$, and the upper limit of the integral in Equation (2.22) can be safely taken to be of the order of a few hundred Fermi.

\section{Applications}

\subsection{Application to the Goldhaber-Teller model of the giant dipole excitation}

The Goldhaber-Teller \cite{GT} model is based upon a division of the target degrees of freedom into three sets:

${\bf p}'_1,...,{\bf p}'_Z$ locating the protons relative to the proton mass-center,

${\bf n}'_1,...,{\bf n}'_N$ locating the neutrons relative to the neutron mass-center,

${\bf R}$ ~~~~~~~~ locating the proton mass-center relative to the neutron mass-center.

States of the giant resonance excitation are postulated to have the form
\beq
\Psi^{n,\ell}_m~=~\chi({\bf p}'_1,...,{\bf p}'_Z;{\bf n}'_1,...,{\bf n}'_N)
\Phi^{n,\ell}_m({\bf R}).
\eeq
All states have a common spherically symmetric $intrinsic$ state $\chi({\bf p}'_1,....{\bf p}'_Z;{\bf n}'_1,....{\bf n}'_N)$, but
they differ in the relative motion of the neutron and proton mass centers, as specified by the different functions $\Phi^{n,\ell}_m({\bf R})$,
which describe small oscillations with an approximately harmonic restoring potential.

The transition charge and current densities between two states $\Psi^{n,\ell}_m$ and $\Psi^{n',\ell'}_{m'}$ are
\bes
\bea
\rho_{n\ell m, n'\ell' m'}({\bf r})&=&e\int d{\bf p}_1,....,d{\bf p}_Z;d{\bf n}_1,....,d{\bf n}_N \nonumber\\ 
&\times&\sum_{i=1}^Z \delta({\bf r}-{\bf p}_i)\Psi^{*n\ell}_m({\bf p}_1,...,{\bf p}_Z;{\bf n}_1,....,{\bf n}_N)\Psi^{n'\ell'}_{m'}({\bf p}_1,...,{\bf p}_Z;{\bf n}_1,....,{\bf n}_N)
\eea 
\bea
{\bf J}_{n\ell m, n'\ell' m'}({\bf r})~&=&~\frac{e\hbar}{2 m_{\rm p} i}\int d{\bf p}_1,....,d{\bf p}_Z; d{\bf n}_1,....,d{\bf n}_N \nonumber \\
&\times&\sum_{i=1}^Z \delta({\bf r}-{\bf p}_i)
 \left(~\Psi^{*n\ell}_m({\bf p}_1,...,{\bf p}_Z; {\bf n}_1,....,{\bf n}_N)\nabla_{{\bf p}_i}\Psi^{n'\ell'}_{m'}({\bf p}_1,...,{\bf p}_Z;{\bf n}_1,....,{\bf n}_N )\right. \nonumber\\
&-&\left. \Psi^{n'\ell'}_{m'}({\bf p}_1,...{\bf p}_Z;{\bf n}_1,....,{\bf n}_N)\nabla_{{\bf p}_i}\Psi^{*n\ell}_m({\bf p}_1,...,{\bf p}_Z; {\bf n}_1,....,{\bf n}_N)\right),
\eea
\ees
where $m_{\rm p}$ represents the proton mass. The vectors ${\bf p}_1,...{\bf p}_Z;{\bf n}_1,....,{\bf n}_N$ locate the target protons and neutrons relative to the target mass center.
They are related to the ${\bf p}'_1,....{\bf p}'_Z;{\bf n}'_1,....{\bf n}'_N$ defined previously by
\bes
\beq
{\bf p}_i~=~f{\bf R}+{\bf p}'_i
\eeq
\beq
{\bf n}_j~=~(f-1){\bf R}+{\bf n}'_j,
\eeq
\ees
where $f~\equiv~\frac{N}{N+Z}$ (see Figure 1).

If we use (3.3) and (3.1) in (3.2a), we get
\begin{eqnarray*}
\rho_{n\ell m, n'\ell' m'}({\bf r})&=&e\int d{\bf R}d{\bf p}'_1,....d{\bf p}'_Z,
d{\bf n}'_1,....d{\bf n}'_N
|\chi_k({\bf p}'_1,...,{\bf p}'_Z;{\bf n}'_1,...,{\bf n}'_N)|^2\\
&\times&\Phi^{*n\ell}_m({\bf R})\Phi^{n'\ell'}_{m'}({\bf R})
\sum_{i=1}^Z\delta({\bf r}-f{\bf R}-{\bf p}'_i)
\end{eqnarray*}
The structure of this expression suggests we define an $intrinsic$ density function
\beq
\rho_{\rm int}({\bf s})\equiv e\int d{\bf p}'_1,....d{\bf p}'_Z;d{\bf n}'_1,....d{\bf n}'_N |\chi({\bf p}'_1,....{\bf p}'_Z;{\bf n}'_1,....{\bf n}'_N)|^2 \sum_{i=1}^Z\delta({\bf s}-{\bf p}'_i),
\eeq
in terms of which $\rho_{n\ell m, n'\ell' m'}({\bf r})$ and ${\bf J}_{n\ell m, n'\ell' m'}({\bf r})$ can be written
\bes
\beq
\rho_{n\ell m, n'\ell' m'}({\bf r})~ =~ \int d^3R~\rho_{\rm int}({\bf r}-f{\bf R})\Phi^{*n\ell}_m({\bf R})\Phi^{n'\ell'}_{m'}({\bf R})
\eeq
\beq
{\bf J}_{n\ell m, n'\ell' m'}({\bf r})~ =~ \frac{\hbar}{2 Zm_{\rm p} i}\int d^3R~\rho_{\rm int}({\bf r}-f{\bf R})
\left(\Phi^{*n\ell}_m({\bf R})\nabla_{\bf R}\Phi^{n'\ell'}_{m'}({\bf R})-\Phi^{n'\ell'}_m({\bf R})\nabla_{\bf R}\Phi^{*n\ell}_{m}({\bf R})\right)
\eeq
\ees
The physical significance of $\rho_{\rm int}({\bf s})$ is seen from Figure 1 to be the charge density defined relative to the mass center of the proton distribution. 

The amplitude of the oscillation in ${\bf R}$ is determined by the parameter $\nu\equiv M\omega/\hbar$, where $M$ is the reduced mass associated with the relative oscillation of the proton and neutron mass centers, and $\hbar\omega$ is the characteristic energy of the oscillation. For the giant dipole resonance, the oscillation energy is approximately
$$
\hbar\omega\simeq79 A^{-1/3}~{\rm MeV}
$$
(\cite{BM}). Thus
$$
\nu=\frac{NZ}{A}m_{\rm p}\frac{\omega}{\hbar}=\frac{NZ}{A}m_{\rm p}c^2\frac{\hbar\omega}{(\hbar c)^2}\simeq\frac{NZ}{A}\times 939~ {\rm MeV} 
\times\frac{79 A^{-1/3}~ {\rm MeV}}{(197.3~ {\rm MeV~ fm})^2} \simeq 1.91\frac{NZ}{A^{4/3}}~ {\rm fm}^{-2}
$$
The amplitude of the oscillation of the proton mass center is then
\beq
f\times \frac{1}{\sqrt{\nu}}=\frac{N}{A}\times \frac{1}{\sqrt{\nu}}\simeq\frac{0.724}{A^{1/3}}\sqrt{\frac{N}{Z}}~ {\rm fm}.
\eeq
If this distance is small compared to the distance over which $\rho_{\rm int}({\bf r})$ changes by an appreciable fraction of itself, such as the thickness of the nuclear surface, then it should be useful to do a Taylor expansion of $\rho_{\rm int}({\bf r}-f{\bf R})$ about ${\bf R}=0$:
\beq
\rho_{\rm int}({\bf r}-f{\bf R})\simeq\rho_{\rm int}({\bf r})-f{\bf R}\cdot\nabla_{\bf r}\rho_{\rm int}({\bf r})+~ \cdots
\eeq
This enables us to approximate (3.5a and b) by
\bes
\bea
\rho_{n\ell m, n'\ell' m'}({\bf r})&\simeq&~ \int d^3R [\rho_{\rm int}({\bf r})-f{\bf R}\cdot\nabla_{\bf r}\rho_{\rm int}({\bf r})]\Phi^{*n\ell}_m({\bf R})\Phi^{n'\ell'}_{m'}({\bf R})\nonumber\\
&=&\delta_{n,n'}\delta_{\ell,\ell'}\delta_{m,m'}\rho_{\rm int}({\bf r})-f\nabla_{\bf r}\rho_{\rm int}({\bf r})\cdot\int\Phi^{*n\ell}_m({\bf R}){\bf R}\Phi^{n'\ell'}_{m'}({\bf R})d^3R
\eea
\bea
{\bf J}_{n\ell m, n'\ell' m'}({\bf r}) &\simeq& \frac{\hbar}{2 Zm_{\rm p} i}\int d^3R \rho_{\rm int}({\bf r})
[\Phi^{*n\ell}_m({\bf R})\nabla_{\bf R}\Phi^{n'\ell'}_{m'}({\bf R})-\Phi^{n'\ell'}_{m'}({\bf R})\nabla_{\bf R}\Phi^{*n\ell}_{m}({\bf  R})]\nonumber\\
&=&\frac{\rho_{\rm int}({\bf r})}{Zm_{\rm p}}\int \Phi^{*n\ell}_m({\bf R}){\bf P}\Phi^{n'\ell'}_{m'}({\bf R})d^3R
\eea
\ees
If we combine these expressions with (2.4), we get 
\bea
<\Phi^{n\ell}_{m}|V(t)|\Phi^{n'\ell'}_{m'}>&=&\delta_{n,n'}\delta_{\ell,\ell'}\delta_{m,m'}\int d^3r \varphi_{_C}^{\rm ret}({\bf r},t)
\rho_{\rm int}({\bf r})\nonumber\\
&-&f\int d^3r \varphi_{_C}^{\rm ret}({\bf r},t)\nabla_{\bf r}\rho_{\rm int}({\bf r})\cdot\int \Phi^{*n\ell}_m({\bf R}){\bf R}\Phi^{n'\ell'}_{m'}({\bf R})d^3R\nonumber\\
&-&\frac{1}{Zm_{\rm p}c}\int d^3r {\bf A}_{_C}^{\rm ret}({\bf r},t)\rho_{\rm int}({\bf r})~\cdot~\int \Phi^{*n\ell}_m({\bf R}){\bf P}\Phi^{n'\ell'}_{m'}({\bf R})d^3R
\eea
The first term is a monopole integral, independent of the internal degrees of freedom of the target. Its effect can be absorbed into a time-dependent
phase factor multiplying the wave function, and we ignore it in the following discussion. We then see that the remaining terms in (3.9) are of the form (2.5), with 
\bes
\bea
{\bf F}(t)&=&\frac{N}{A}\int d^3r \varphi_{_C}^{\rm ret}({\bf r},t)\nabla_{\bf r}\rho_{\rm int}({\bf r})\\
{\bf G}(t)&=&\frac{v}{Zm_{\rm p}c}\int d^3r {\bf A}_{_C}^{\rm ret}({\bf r},t)\rho_{\rm int}({\bf r})\hat{z}~ .
\eea
\ees

To estimate the validity of the approximation (3.7), let us consider the particular example of a $^{40}$Ca
target. Then Equation (3.6) yields $f/\sqrt{\nu}\simeq 0.724/40^{1/3}~ {\rm fm}\simeq 0.212~{\rm fm}$ for the amplitude of the oscillation of the proton mass center. Since the proton charge density is approximately constant from its center out to the surface region, whose thickness is approximately 1 fm, we see that the amplitude of the GDR oscillation is indeed small compared to the distance over which the charge density changes by an appreciable fraction of itself. Thus we would expect that Equation (3.7) is a reasonable first step in the analysis of GDR data.

We have not used the term ``electric dipole approximation" to label (3.7), since that term usually implies a comparison of a wavelength with the size of a charge distribution. This is not the comparison we need to justify (3.7).

We now distinguish between regimes in which the projectile is relativistic or non-relativistic:
\subsubsection {Relativistic projectiles}
In this approximation \cite{WA}, the projectile linear momentum is very large compared to the transverse impulse the projectile receives as it moves past the target. Then the trajectory of the projectile can be approximated by a straight line, along which the projectile moves with constant speed $v$. We choose our axes with the target center at the origin, the projectile trajectory in the ${\bf{\hat y}}-{\bf{\hat z}}$ plane, a constant distance $b$ from the ${\bf{\hat z}}$ axis, and the projectile moving in the ${\bf{\hat z}}$ direction. Then $\varphi_{_C}^{\rm ret}({\bf r},t), {\bf A}^{\rm ret}({\bf r},t)$ are the Lienard-Wiechert potentials \cite{JA}
\bes
\bea
\varphi_{_C}^{\rm ret}({\bf r},t)&=&\frac{\gamma Z_{\rm P}e}{\sqrt{x^2+(y-b)^2+\gamma^2(z-vt)^2}}\\
{\bf A}^{\rm ret}({\bf r},t)&=&\frac{v}{c}\varphi_{_C}^{\rm ret}({\bf r},t){\bf{\hat z}}
\eea
\ees
We can now use (3.10) and (2.8b) to calculate the $\alpha_x,\alpha_y,\alpha_z$ needed in (2.20):
\bea
\alp&=&i\int_{-\infty}^\infty dt'\left[\frac{{\bf F}(t')}{\sqrt{2M\hbar\omega}}+i\sqrt{\frac{M\omega}{2\hbar}}{\bf G}(t')\right]e^{i\omega t'}\nonumber\\
&=&i\sqrt{\frac{\hbar}{2M\omega}}\frac{N}{A}\int d^3r \varphi_{_C}^{\rm ret}({\bf r},\omega)\nabla_{\bf r}\rho_{\rm int}({\bf r})+
\sqrt{\frac{M\omega\hbar}{2}}\frac{v}{Zm_{\rm p}c^2}\int d^3r \varphi_{_C}^{\rm ret}({\bf r},\omega)\rho_{\rm int}({\bf r})\hat{z}
\eea
where
$$
\varphi_{_C}^{\rm ret}({\bf r},\omega)~\equiv~ \int^{\infty}_{-\infty}\frac{dt'}{\hbar}\varphi_{_C}^{\rm ret}({\bf r},t')e^{i\omega t'}
$$
is the ``on-shell" Fourier transform of the retarded potential (3.11a). A convenient multipole expansion of this function is \cite{WA,BZ}
\bes
\beq
\varphi_{_C}^{\rm ret}({\bf r},\omega)~ =~ \frac{2Z_{\rm p}e}{\hbar v}\sum_m e^{-im\frac{\pi}{2}}K_m\left(\frac{|\omega| b}{\gamma v}\right)
\sum_{\lambda=|m|}^{\infty}{\cal G}_{\lambda,m}j_\lambda\left(\frac{|\omega|}{c}r\right)Y_{m}^{\lambda}(\hat{r}),
\eeq
where the coefficients ${\cal G}_{\lambda,m}$ are defined by 
\beq
{\cal G}_{\lambda,m}\equiv\frac{i^{\lambda+m}}{(2\gamma)^m}\left(\frac{\omega}{|\omega|}\right)^{\lambda-m}(\frac{c}{v})^\lambda
\sqrt{4\pi(2\lambda+1)(\lambda-m)!(\lambda+m)!}\sum_n\frac{1}{(2\gamma)^{2n}(n+m)!n!(\lambda-m-2n)!}
\eeq
\ees
We assume that the target proton charge distribution is spherically symmetric, so $\rho_{\rm int}({\bf r})=\rho_{\rm int}(r)$. Then
\bes
\beq
\int \sin\theta d\theta d\varphi Y_{m}^{\lambda}(\theta,\varphi)\nabla \rho_{\rm int}(r)=\delta_{\lambda,1}\sqrt{\frac{4\pi}{3}}\rho'_{\rm int}(r)\times q(m)
\eeq
where $q(1)=-\frac{\hat{x}+i\hat{y}}{\sqrt{2}}$, $q(0)=\hat{z}$ and $q(-1)=\frac{\hat{x}-i\hat{y}}{\sqrt{2}}$, and
\beq
\int \sin\theta d\theta d\varphi Y_{m}^{\lambda}(\theta,\varphi)\rho_{\rm int}(r)=\delta_{\lambda,0}\delta_{m,0}\sqrt{4\pi}\rho_{\rm int}(r)
\eeq
\ees
If (3.11) is substituted into (3.10), and (3.14) is used to evaluate the angular integrals, the result is
\bes
\begin{eqnarray}
\alpha_x&=&0\\
\alpha_y &=& -i\pi \sqrt{\frac{16N\omega}{m_{\rm p}ZA\hbar}}\frac{Z_{\rm p}e^2}{\gamma v^2}
K_1\left(\frac{\omega b}{\gamma v}\right)\int_0^\infty 
j_0\left(\frac{\omega}{c}r\right)\rho_{\rm int}(r)r^2dr \\
\alpha_z&=& \pi \sqrt{\frac{32N\omega}{m_{\rm p}ZA\hbar}}\frac{Z_{\rm p}e^2}{\gamma^2 v^2}
K_0\left(\frac{\omega b}{\gamma v}\right)\int_0^\infty 
j_0\left(\frac{\omega}{c}r\right)\rho_{\rm int}(r)r^2dr 
\end{eqnarray}
\ees
Use has been made of the relation
$$
\int_{0}^\infty j_1\left(\frac{\omega}{c}r\right)\rho'_{\rm int}(r)r^2 dr =~ -\frac{\omega}{c}
\int_{0}^\infty j_0\left(\frac{\omega}{c}r\right)\rho_{\rm int}(r)r^2 dr
$$
This result is gauge invariant, since it involves only on-shell Fourier components of the interaction.
\subsubsection{Non-relativistic projectiles}
Here Newtonian mechanics is used to obtain a hyperbolic trajectory for the projectile \cite{AW}. Following the conventions used in the previous Section, we choose the coordinate axes so that the target center is at the origin, the trajectory is in the ${\bf{\hat y}}-{\bf{\hat z}}$ plane, reflection-symmetric across the ${\bf {\hat y}}$ axis, with the projectile moving in the direction of increasing $z$. If ${\bf r'}(t)$ locates the projectile center at time $t$, then the scalar and vector potentials are
\bes
\bea
\varphi({\bf r},t)=\frac{Z_{\rm P}e}
{|{\bf r - r'}(t)|}\\
{\bf A(r},t)=\frac{{\bf v}(t)}{c}\varphi({\bf r},t)
\eea
\ees
Here ${\bf v}(t)=d{\bf r'}/{dt}$ is the projectile velocity at time t.

The hyperbolic trajectory is characterized by the lengths of its semi-transverse and semi-conjugate axes, $a$ and $b$, related to the asymptotic kinetic energy $E$ and the angular momentum $\ell$ by
\beq
a=\frac{Z_{{\rm P}}Z_{{\rm T}}e^2}{2E}~~~~~~~b=\frac{\ell}{2mE}~~{\rm (impact~parameter)}.
\eeq
The {\it eccentricity} of the trajectory is $\epsilon \equiv \sqrt{1+(b/a)^2}$. The {\it adiabaticity} parameter is $\xi \equiv a\omega/v$, with $v$ the asymptotic relative speed ($v=\sqrt{2E/m}$). Then ${\bf F}(t)$and ${\bf G}(t)$ of eqn.(3.10) are replaced by
\bes
\bea
{\bf F}(t)&=&\frac{N_{{\rm T}}}{A_{{\rm T}}}\int d^3r \frac{Z_{{\rm P}}e}{|{\bf r-r'}(t)|}\rho'_{{\rm int}}(r) {\bf {\hat r}}\nonumber\\
&=&-\frac{N_{{\rm T}}}{A_{{\rm T}}}Z_{{\rm P}}Z_{{\rm T}}e^2
\frac{\left({\bf {\hat x}}~x'(t)+{\bf {\hat y}}~y'(t)+{\bf {\hat z}}~z'(t)\right)}{[r'(t)]^3}\\
{\bf G}(t)&=&\frac{Z_{{\rm P}}e}{Z_{{\rm T}}m_p c^2}\int d^3 r \rho_{{\rm int}}(r)~\left[~\frac {{\bf {\hat x}}~{\dot x'}(t)~+~{\bf {\hat y}}~{\dot y'}(t)~+~{\bf {\hat z}}~{\dot z'}(t)}{r'(t)}~\right]\nonumber\\
&=&\frac{Z_{{\rm P}}e^2}{m_p c^2}\left[~\frac {{\bf {\hat x}}~{\dot x'}(t)~+~{\bf {\hat y}}~{\dot y'}(t)~+~{\bf {\hat z}}~{\dot z'}(t)}{r'(t)}~\right]
\eea
\ees
The time integrals corresponding to eqn. (3.12) can be performed using the methods given by Alder and Winther \cite{AW}. Some details are given in Appendix D. The result is
\bes
\bea
\alpha_x &=& 0\\
\alpha_y &=& i Z_{{\rm P}}Z_{{\rm T}}\frac{e^2}{\hbar c}\sqrt{\frac{2N_{{\rm T}}\hbar \omega}{Z_{{\rm T}}A_{{\rm T}}m_pc^2}}~e^{-\xi\frac{\pi}{2}}\left[\frac{\gamma^2}{\gamma^2-1}I_1-I_2\right]\\
\alpha_z &=& Z_{{\rm P}}Z_{{\rm T}}\frac{e^2}{\hbar c}\sqrt{\frac{2N_{{\rm T}}\hbar \omega}{Z_{{\rm T}}A_{{\rm T}}m_pc^2}}~e^{-\xi\frac{\pi}{2}}\frac{b}{a}\left[\frac{\gamma^2}{\gamma^2-1}\frac{I_3}{\epsilon}-I_4\right],
\eea
\ees
where
\bes
\bea 
I_1&\equiv& -\int_0^\infty e^{-\xi \epsilon \cosh(w)}\cosh(w)\cos(\xi w) dw\\
I_2&\equiv& \int_0^\infty e^{-\xi \epsilon \cosh(w)}\cosh(w)\left[~\frac{\cos(\xi w)+\epsilon \sinh(w)\sin(\xi w)}{1+(\epsilon \sinh(w))^2}\right] dw\\
I_3&\equiv& \int_0^\infty e^{-\xi \epsilon \cosh(w)}\cos(\xi w) dw\\
I_4&\equiv& \int_0^\infty e^{-\xi \epsilon \sinh(w)}\sinh(w)\left[~\frac{-\sin(\xi w)+\epsilon \sinh(w)\cos(\xi w)}{1+(\epsilon \sinh(w))^2}\right] dw.
\eea
\ees
These integrals are easily calculated numerically. Note that in this non-relativistic approximation, only the total target charge $Z_{{\rm T}}$ is relevant, not the radial charge density distribution.
\subsubsection{Overlap between relativistic and non-relativistic regions.}
It is of some interest to consider the transition between the regions of applicability of the relativistic and non-relativistic approaches to Coulomb excitation. The principal approximation limiting the validity of the relativistic approach, as the bombarding energy decreases, is the assumption that the projectile moves along a straight-line trajectory at constant speed. Actually, at low speeds the projectile is deflected by the Coulomb field of the target, with a classical scattering angle of $\theta=2{\rm ArcCot}\left(\frac{b}{a}\right)=2{\rm ArcCot}\left(\frac{2Eb}{Z_{{\rm P}}Z_{{\rm T}}e^2}\right)$. 
As an example, consider $^{208}$Pb projectiles Coulomb deflected by a $^{40}$Ca target. Then the deflection angle would be less than $1^{\circ}$ if $Eb>\frac{82 \times 20 \times 1.44}{2}\times {\rm Cot}(.5^{\circ})=1.35\times 10^{5}$ MeV fm. For a Coulomb-dominated trajectory, $b\geq 12$ fm, so we will not encounter trajectory deflections more than 1$^{\circ}$ if $E>1.13\times10^4$ MeV, or a projectile energy of $1.13\times10^4/208\sim 54$ MeV per nucleon. Moreover, for an assumed straight-line trajectory with $b=12$ fm, the decrease in projectile speed as the Pb projectile moves past the Ca target is less than 1\%. Thus we would not expect the deviations from a constant velocity orbit to cause significant problems for the relativistic approximation in the $^{208}$Pb-$^{40}$Ca system as long as the projectile kinetic energy per nucleon exceeds 54 MeV.

Another aspect of the relativistic approximation is the neglect of the recoil of the target as the projectile moves past it. Winther and Alder \cite{WA} have shown that this may be corrected to some extent by replacing the relativistic adiabaticity parameter $\frac{\omega b}{\gamma v}$ in eqn (3.15) by 
$$
\frac{\omega}{\gamma v}\left(b+\frac{\pi a}{2 \gamma}\right).
$$
This correction is applied in the results to be shown below.

The most serious error introduced when the non-relativistic approximation is applied to fast projectiles is the omission of the relativistic sharpening of the electromagnetic pulse experienced by the target. It is this sharpening that causes the adiabaticity parameter to be $\frac{\omega b}{\gamma v}$, rather than $\frac{\omega b}{v}$. The presence of the $1/\gamma$ factor increases the amplitudes of high-frequency components in the pulse, which are necessary for the population of a high-energy excitation mode such as the GDR. Thus we might expect the non-relativistic Coulomb excitation formalism to underestimate the GDR excitation cross-section as the projectile kinetic energy increases into the relativistic region.

Figure 2 shows a comparison of the results of relativistic and non-relativistic calculations for the excitation of the one-phonon GDR level in $^{40}$Ca, as a result of $^{208}$Pb-induced Coulomb excitation. The curve labelled ``relativistic" is certainly applicable at the high-bombarding-energy end of the energy range. The above qualitative considerations indicate that it may also be valid down to bombarding energies per nucleon of about 50 MeV. This inference is supported by the fact that at 50 MeV per nucleon, it agrees with the non-relativistic result, which should be valid at this lower energy. Thus there is good reason to believe that, in this case, the relativistic result is valid down to about 50 MeV per nucleon. At this energy the magnitude of the cross-section is so low that detection of the GDR would be difficult. Thus the relativistic theory seems to be valid over the entire useful energy range.

Figure 2 also shows that the difference between the predictions of the relativistic and non-relativistic theories, as the bombarding energy increases, is in the expected direction.

\subsection{Application to the Steinwedel-Jensen model of the giant dipole excitation}
This is a model \cite{SJ} in which protons and neutrons oscillate relative to each other, not in the bulk relative motion of the Goldhaber-Teller model \cite{GT}, but
in local isovector fluctuations. 
Our version of this model follows the presentation of Greiner and Maruhn \cite{GR}. We present only the version of the theory in which the projectile is relativistic, because we have seen in the last section that this theory is applicable over the entire energy range for which multiphonon GDR levels are observable. 

Let ${\bf s}_{_N}({\bf r},t)$ and ${\bf s}_{_P}({\bf r},t)$ be neutron and proton displacement fields, i.e. when a fluctuation occurs a
neutron that was in equilibrium at ${\bf r}$ moves to ${\bf r}+{\bf s}_{_N}({\bf r},t)$. Because of the isovector nature of the 
fluctuation, ${\bf s}_{_N}({\bf r},t)$ and ${\bf s}_{_P}({\bf r},t)$ can be expressed in terms of the single vector 
field ${\bf s}({\bf r},t)$:
\bes
\bea
{\bf s}_{_P}({\bf r},t)&=&\frac{N}{A}{\bf s}({\bf r},t)\\
{\bf s}_{_N}({\bf r},t)&=&-\frac{Z}{A}{\bf s}({\bf r},t)~.
\eea
\ees
The corresponding velocity fields are the time derivatives:
\bes
\bea
{\bf v}_{_P}({\bf r},t)&=&\frac{\partial {\bf s}_{_P}({\bf r},t)}{\partial t}~=~\frac{N}{A}\frac{\partial {\bf s}({\bf r},t)}{\partial t}
~\equiv~\frac{N}{A}{\bf v}({\bf r},t)\\
{\bf v}_{_N}({\bf r},t)&=&\frac{\partial {\bf s}_{_N}({\bf r},t)}{\partial t}~=~-\frac{Z}{A}\frac{\partial {\bf s}({\bf r},t)}{\partial t}
~\equiv~-\frac{Z}{A}{\bf v}({\bf r},t)
\eea
\ees

Let $n_{_0}$ be the equilibrium nucleon number density, so that $\frac{Z}{A}n_{_0}$ and $\frac{N}{A}n_{_0}$ are the equilibrium proton
and neutron number densities. As a result of the displacements (3.21), these number densities become
\bes
\bea
n_{_P}({\bf r},t)&=&\frac{Z}{A}n_{_0}\left(1-\nabla\cdot{\bf s}_{_P}({\bf r},t)\right)~=~\frac{Z}{A}n_{_0}~+~\eta({\bf r},t) \\
n_{_N}({\bf r},t)&=&\frac{N}{A}n_{_0}\left(1-\nabla\cdot{\bf s}_{_N}({\bf r},t)\right)~=~\frac{N}{A}n_{_0}~-~\eta({\bf r},t)
\eea
\ees
where the isovector density fluctuation $\eta({\bf r},t)$ is related to the isovector displacement field ${\bf s}({\bf r},t)$ by
\beq
\eta({\bf r},t)~=~-\frac{NZ}{A^2}n_{_0}\nabla\cdot{\bf s}({\bf r},t)
\eeq

For small fluctuations, the kinetic and potential energy densities are given by
\bes
\bea
{\cal T}({\bf r},t)&=&\frac{1}{2}[n_{_P}({\bf r},t)({\bf v}_{_P}({\bf r},t))^2+n_{_N}({\bf r},t)({\bf v}_{_N}({\bf r},t))^2]\nonumber\\
&\simeq&n_0\frac{m_{{\rm p}}}{2}\frac{ZN}{A^2}\left[\left(\frac{\partial s_x({\bf r},t)}{\partial t}\right)^2 \right.
\left. +\left(\frac{\partial s_y({\bf r},t)}{\partial t}\right)^2+ \left(\frac{\partial s_z({\bf r},t)}{\partial t}\right)^2\right]\\
{\cal U}({\bf r},t)&=&4\frac{a_s}{n_0}\left(\eta({\bf r},t)\right)^2
~=~ 4\frac{a_s}{n_0}\left(\frac{NZ}{A^2}\right)^2\left(\frac{\partial s_x({\bf r},t)}{\partial x}+\right.
\left.\frac{\partial s_y({\bf r},t)}{\partial y}+\frac{\partial s_z({\bf r},t)}{\partial z}\right)^2
\eea
\ees
Here $a_s$ ($\simeq 23$ MeV) is the symmetry energy parameter. If the Lagrange equations of motion 
$$
\frac{\partial}{\partial t}\left(\frac{\partial {\cal L}}{\partial \frac{\partial s_\alpha}{\partial t}}\right)+
\frac{\partial}{\partial x}\left(\frac{\partial {\cal L}}{\partial \frac{\partial s_\alpha}{\partial x}}\right)+
\frac{\partial}{\partial y}\left(\frac{\partial {\cal L}}{\partial \frac{\partial s_\alpha}{\partial y}}\right)+
\frac{\partial}{\partial z}\left(\frac{\partial {\cal L}}{\partial \frac{\partial s_\alpha}{\partial z}}\right)
~=~ \frac{\partial {\cal L}}{\partial s_\alpha}
$$
are applied to the Lagrangian density ${\cal L}={\cal T}-{\cal U}$, the result is
\beq
n_0m_{{\rm p}}\frac{\partial^2 {\bf s}}{\partial t^2}\kk=\kk n_0m_{{\rm p}}\frac{\partial {\bf v}}{\partial t}\kk=-\kk 8a_s\nabla\eta
\eeq

Number conservation for the protons and neutrons
$$
\frac{\partial n_{\rm P}}{\partial t}=-\nabla\cdot(n_{\rm P}{\bf v}_{\rm P})=-\frac{Z}{A}n_0\nabla\cdot{\bf v}_{\rm P}, \kko
\frac{\partial n_{\rm N}}{\partial t}=-\nabla\cdot(n_{\rm N}{\bf v}_{\rm N})=-\frac{Z}{A}n_0\nabla\cdot{\bf v}_{\rm N}
$$
can be expressed with the help of (3.22) and (3.23) as
\beq
\frac{\partial\eta}{\partial t}\kk=\kk-n_0\frac{NZ}{A^2}\nabla\cdot{\bf v}
\eeq
Combining (3.26) and (3.27) leads to the d'Alambertian equation 
\bes 
\beq
\frac{\partial^2 \eta}{\partial t^2}\kk=\kk u^2\kk\nabla^2\eta
\eeq
with
\beq
u\kk=\kk\sqrt{\frac{8a_s}{m_{{\rm p}}}\frac{NZ}{A^2}}.
\eeq
\ees

The Steinwedel-Jensen model of giant dipole excitations \cite{SJ} is based on solutions of (3.28a) of the form 
\bes
\beq
\eta({\bf r},t)\kk=\kk \frac{NZ}{A^2}n_0 \frac{j_1(\frac{\omega}{u}r)}{r}{\bf r\cdot a}(t)
\eeq
where the time-dependent amplitude ${\bf a}(t)$ exhibits harmonic oscillations with frequency $\omega$:
\beq
\frac{d^2 {\bf a}(t)}{dt^2}~ =~ -\omega^2{\bf a}(t)
\eeq
\ees
To show that (3.29) satisfies (3.28), we observe that $\frac{{\bf r}}{r}\cdot {\bf a}(t)$ is a linear combination of spherical harmonics
$Y^1_m(\hat{r})$, and that
$$
\nabla^2 j_1\left(\frac{\omega}{u}r\right)Y^1_m(\hat{r})~=~ -\left(\frac{\omega}{u}\right)^2 j_1\left(\frac{\omega}{u}r\right)Y^1_m(\hat{r}).
$$
The irrotational displacement field consistent with (3.29) and (3.24) is
\beq
{\bf s}({\bf r},t)\kk=\kk\frac{u^2}{\omega^2}\kk \nabla \left(\frac{j_1(\frac{\omega}{u}r)}{r}{\bf r\cdot a}(t)\right)
\eeq

The boundary condition that determines $\omega$ is the requirement that the radial component of ${\bf s}({\bf r},t)$ should vanish at the nuclear surface. If this were not true, fluctuations would imply a complete separation of protons from neutrons on either side of the nuclear surface
which is regarded as unphysical (although it occurs in the Goldhaber-Teller model \cite{GT}). Thus it is required that
$$
0=\hat{r}\cdot{\bf s}({\bf r},t)|_{r=R}=\frac{u^2}{\omega^2}\left|\frac{d}{dr}j_1\left(\frac{\omega}{u}r\right)\right|\hat{r}\cdot{\bf a}(t)+j_1\left(\frac{\omega}{u}R\right) \frac{d}{dr}\left(\hat{r}\cdot{\bf a}(t)\right)
\kk=\kk\frac{u}{\omega}j'_1\left(\frac{\omega}{u}R\right)\hat{r}\cdot{\bf a}(t)
$$
It is therefore required that $\frac{\omega}{u}R$ be a zero of $j'_1\left(\frac{\omega}{u}R\right)$. The lowest zero is $\sim 2.08$, and 
$\omega$ is to be determined by the eigenvalue condition
\beq
\omega~=~2.08\frac{u}{R}
\eeq

So far the picture is one of classical harmonic oscillations of a two-fluid system within a closed spherical volume of radius $R$. 
To quantize this picture, we quantize the ``virtual oscillator'' (3.29b), i.e. we treat $a_x, a_y, a_z$ as the coordinates of a quantum 
harmonic oscillator of frequency $\omega$. The states of the giant dipole excitation would be eigenstates of this virtual oscillator.
To get the size parameter associated with this virtual oscillator we identify its potential energy 
$\frac{1}{2}M\omega^2(a^2_x+ a^2_y+ a^2_z)$
with the volume integral of the potential energy density (3.25b)
\begin{eqnarray*}
\int_{r<R}d^3r~ {\cal U}({\bf r},t)&=&\frac{4a_s}{n_0}\int_{r<R}d^3r(\eta({\bf r},t))^2=4a_sn_0\left(\frac{NZ}{A^2}\right)^2 \int_{r<R}d^3r
\left(\frac{j_1(\frac{\omega}{u}r)}{r}{\bf r}\cdot{\bf a}\right)^2\\
&=&\frac{1}{2}M\omega^2(a^2_x+ a^2_y+ a^2_z)
\end{eqnarray*}
A simple calculation shows that the size parameter of the virtual oscillator is
\beq
\sqrt{\frac{\hbar}{M\omega}}~ =~ \sqrt{\frac{3\hbar\omega^4}{32\pi a_sn_0\left(\frac{NZ}{A^2}\right)^2
u^3\int_0^{2.08}(qj_1(q))^2dq}}~=~\sqrt\frac{\hbar\omega(2.08A)^3}{8a_s\times 0.462} .
\eeq

Now the charge and current densities can be expressed in terms of the operators $a_x, a_y, a_z$ and the conjugate momenta $P_x,P_y,P_z$:
\bes
\bea
\rho({\bf r})&=&en_{\rm P}({\bf r})~=~ \frac{Z}{A}n_0+e\eta({\bf r})\nonumber\\  
&=&e\frac{Z}{A}n_0+e\frac{NZ}{A^2}n_0   \frac{j_1(\frac{\omega}{u}r)}{r}{\bf r\cdot a}
\eea
\bea
{\bf J}({\bf r})&=&en_{\rm P}({\bf r}){\bf v}_{\rm P}({\bf r})~\simeq~e\frac{ZN}{A^2}n_0{\bf v}({\bf r})=
e\frac{ZN}{A^2}n_0\frac{\partial}{\partial t}{\bf s}({\bf r},t)\nonumber\\
&=& e\frac{ZN}{A^2}n_0\frac{u^2}{\omega^2}\nabla\left(\frac{j_1(\frac{\omega}{u}r)}{r}{\bf r}\cdot\dot{{\bf a}}\right)
= e\frac{ZN}{A^2}n_0\frac{u^2}{M\omega^2}\nabla\left(\frac{j_1(\frac{\omega}{u}r)}{r}{\bf r}\cdot{\bf P}\right)
\eea
\ees
If these charge and current densities are substituted into (2.4), we get an expression of the form (2.5) (with ${\bf R}$ of (2.5) replaced 
by ${\bf a}$ here), from which we can extract
\bes
\begin{eqnarray}
{\bf F}(t)&=&-\frac{eNZ}{A^2}n_0\int d^3r \varphi_c^{\rm ret}({\bf r},t)~\frac{j_1(\frac{\omega}{u}r)}{r}{\bf r}\\
{\bf G}(t)&=&\frac{v}{Mc^2}\frac{eNZ}{A^2} n_0(\frac{u}{\omega})^2\int d^3r \varphi_c^{\rm ret}({\bf r},t)\nabla 
\left(\frac{j_1(\frac{\omega}{u}r)}{r}{\bf r}\right)
\end{eqnarray}
\ees
and
\begin{eqnarray*}
\alp&=&i\int_{-\infty}^\infty dt'\left[\frac{{\bf F}(t')}{\sqrt{2M\hbar\omega}}+i\sqrt{\frac{M\omega}{2\hbar}}{\bf G}(t')\right]
e^{i\omega t'}\\
&=&-i\sqrt{\frac{\hbar}{2M\omega}}\frac{eNZ}{A^2}n_0\int d^3r \varphi_c^{\rm ret}({\bf r},\omega)\frac{j_1(\frac{\omega}{u}r)}{r}{\bf r}\\
&-&\sqrt{\frac{M\hbar\omega}{2}}\frac{v}{Mc^2}
\frac{eZN}{A^2}n_0 (\frac{u}{\omega})^2\int d^3r\varphi_c^{\rm ret}({\bf r},\omega)\frac{\partial}{\partial z} 
\left(\frac{j_1(\frac{\omega}{u}r)}{r}{\bf r}\right)
\end{eqnarray*}
With the help of (3.13), we find
\bes
\bea
\alpha_x&=&0\\
\alpha_y &=&-i\frac{8\pi NZ}{A^2} \frac{Z_{\rm p}e^2n_0}{\hbar\gamma v}\sqrt{\frac{\hbar}{2M\omega}}K_1(\frac{\omega b}{\gamma v})\nonumber\\
&&\times\left[\frac{c}{v}\int_0^Rr^2dr j_1(\frac{\omega}{u}r) j_1(\frac{\omega}{c}r)-\frac{u}{v}\int_0^Rr^2dr j_2(\frac{\omega}{u}r)\left. 
\right.j_2(\frac{\omega}{c}r)\right]
\eea
\bea 
\alpha_z& =&-\frac{8\pi NZ}{A^2}\frac{Z_{\rm p}e^2n_0}{\hbar v}\sqrt{\frac{\hbar}{2M\omega}} K_0(\frac{\omega b}{\gamma v})
\times\left[\frac{1}{3}\frac{uv}{c^2} \int_0^R j_0(\frac{\omega}{u}r)j_0(\frac{\omega}{c}r)r^2dr  \right. \nonumber\\
&-&\left.\frac{c}{v}\int_0^Rr^2drj_1(\frac{\omega}{u}r) j_1(\frac{\omega}{c}r)\right.
\left. +\frac{2}{3}\frac{u}{v}\left(1+\frac{1}{2\gamma^2}\right)\int_0^Rr^2dr \right.
\left.j_2(\frac{\omega}{u}r) j_2(\frac{\omega}{c}r)\right] 
\eea
\ees

\section{High bombarding-energy limit}

According to Equation (3.15), the high-bombarding-energy behavior of $\alpha_y$ and $\alpha_z$ in the Goldhaber-Teller model \cite{GT} is determined by
\bea
\alpha_y&\stackrel{\gamma\rightarrow \infty}{\longrightarrow}&-i\pi\sqrt{\frac{32N}{m_{{\rm p}}ZA\hbar\omega}}~\frac{Z_{{\rm p}}e^2}{cb}~\int_0^\infty j_0\left(\frac{\omega}{c}r\right)\rho_{{\rm int}}(r)r^2 dr \\
\alpha_z&\stackrel{\gamma\rightarrow \infty}{\longrightarrow}&\pi\sqrt{\frac{32N\omega}{m_{{\rm p}}ZA\hbar}}~\frac{Z_{{\rm p}}e^2}{\gamma^2 b^2}~{\rm Ln}\left(\frac{\gamma c}{\omega b}\right)~\int_0^\infty j_0\left(\frac{\omega}{c}r\right)\rho_{{\rm int}}(r)r^2 dr 
\eea
Thus $\alpha_y$ has a finite high-bombarding-energy limit, whereas $\alpha_z$ approaches zero as ${{\rm Ln}}(\gamma)/\gamma^2$. The corresponding expressions for the Steinwedel-Jensen model \cite{SJ} are obtained from Equations (3.35):
\bes
\beq
\alpha_y\stackrel{\gamma\rightarrow \infty}{\longrightarrow} i\frac{8\pi NZ}{A^2}\frac{Z_{{\rm p}}e^2 c^2 u^2 n_0}{\hbar \omega^5 b R}~\sqrt{\frac{\hbar}{2M\omega}}~j_1\left(\frac{\omega R}{c}\right)\left[4\frac{\omega R}{u}\cos\left(\frac{\omega R}{u}\right)+\left(\left(\frac{\omega R}{u}\right)^2-4\right)\sin\left(\frac{\omega R}{u}\right)\right]
\eeq
$$
\alpha_z\stackrel{\gamma\rightarrow \infty}{\longrightarrow} -\frac{8\pi NZ}{A^2}\frac{Z_{{\rm p}}e^2 c^2 n_0}{\hbar c}~\sqrt{\frac{\hbar}{2M\omega}}~K_0\left(\frac{\omega b}{\gamma c}\right)j_1\left(\frac{\omega R}{c}\right)\times
$$
\beq
\left[~\frac{u R^2}{\omega}~j_{1}^{'}\left(\frac{\omega R}{u}\right)~+~\frac{1}{2\gamma^2}\frac{u^4}{\omega^4 R}~\left[4\frac{\omega R}{u}\cos\left(\frac{\omega R}{u}\right)+\left(\left(\frac{\omega R}{u}\right)^2-4\right)\sin\left(\frac{\omega R}{u}\right)\right]\right]
\eeq
\ees
The eigenvalue condition that determines $\omega$ is the vanishing of $j_{1}^{'}\left(\frac{\omega R}{u}\right)$. Therefore the surviving quantity in the last line of Equation (4.3b) is proportional to $1/\gamma^2$, and so 
the high-bombarding-energy dependence of $\alpha_z(\infty)$ is given by 
$$
K_0\left(\frac{\omega b}{\gamma c}\right)/\gamma^2~\stackrel{\gamma\rightarrow \infty}{\longrightarrow}~{\rm Ln}(\gamma)/\gamma^2,
$$
as was the case for the Goldhaber-Teller model. The different behaviors of $\alpha_y(\infty)$ and $\alpha_z(\infty)$ are reminiscent of the distinction between transverse and longitudinal ``pulses" in the Fermi-Weizs$\ddot {\rm a}$cker-Williams method of virtual quanta (\cite{JA}, Chapter 15).

We see that in both the Goldhaber-Teller and Steinwedel-Jensen models, at sufficiently high bombarding energy, $\alpha_z$ will become negligible compared to $\alpha_y$. The geometry of the collision also requires that $\alpha_x=0$. In this situation, the argument of the Legendre polynomial in Equation (2.20) approaches $|\alpha_y|^2/|(\alpha_y)^2|=1$, and the excitation probability of a state of specified $n,~\ell$ reduces to the simpler form
\beq
{\cal P}_{n,\ell}\stackrel{\gamma\rightarrow \infty}{\longrightarrow}\frac{2 \ell +1}{(2n)!!(2n+2\ell +1)!!}\left(|\alpha_y|^2\right)^{2n+\ell}~e^{-|\alpha_y|^2}.
\eeq

Figure 3 illustrates the bombarding-energy dependences of $\alpha_y$ and $\alpha_z$ in the two models. The reaction involves $^{208}$Pb projectiles and a $^{40}$Ca target. The approach of $\alpha_y$ to a constant limiting value, while $\alpha_z$ strongly decreases, is evident. It is also clear from Figure 3, that all the cross-sections predicted by the Goldhaber-Teller and Steinwedel-Jensen models will be very similar, since all the cross-sections are determined by the two parameters $\alpha_y$ and $\alpha_z$.

We get a further simplification of Equation (4.4) if we restrict our attention to levels with a given total number of quanta $N=2n+\ell$. Then we can deduce that
\beq
\frac{{\cal P}_{\frac{N-\ell}{2},\ell}}{{\cal P}_{\frac{N-\ell-2}{2},\ell+2}}\stackrel{\gamma\rightarrow \infty}{\longrightarrow}\frac{(2\ell+1)(N+\ell+3)}{(2\ell+5)(N-\ell)}\stackrel{\gamma\rightarrow \infty}{\longrightarrow} \frac{\sigma_{\frac{N-\ell}{2},\ell}}{\sigma_{\frac{N-\ell-2}{2},\ell+2}}
\eeq
The second relation holds because the ratio is independent of $\alpha_y$, and therefore independent of $b$. According to Equation (2.22), if the ratio of excitation probabilities is independent of $b$, that ratio will also be the cross-section ratio.

In the particular case of 2 phonon levels, Equation (4.5) yields $\sigma_{1,0}/\sigma_{0,2}\stackrel{\gamma\rightarrow \infty}{\longrightarrow}1/2$. This is in agreement with the calculation of Bertulani and Baur \cite{BE}.

Figure 4 shows the bombarding energy dependence of the excitation cross-section for levels with four or fewer GDR phonons in a $^{40}$Ca target, when the projectile is $^{208}$Pb. The calculation was done using the Goldhaber-Teller description of the GDR. As was shown in Section III, the predictions based on the Steinwedel-Jensen description would be similar.

At bombarding energies per nucleon near 10 GeV, Figure 4 exhibits cross-section ratios consistent with Equation (4.5). At lower bombarding energies per nucleon, say below 1 GeV, the cross-section ratios are shown by Figure 4 to be quite different. Changing the bombarding energy has a significant effect on the ratio of transverse and longitudinal impulses received by the target. It follows from Equation (2.20) that this can affect the ratios of excitation probabilities of states of different angular momenta. Indeed, we see that as the bombarding energy per nucleon increases from 1 to 2 GeV, the $N=3$ and $N=4$ levels that are most strongly excited change from $\ell=N$ to $\ell=N-2$. This behavior suggests that some interesting changes in the angular distribution of decay products might be observed as the bombarding energy moves through this region.

\appendix

\section{Proof of the polynomial identity (2.14)}
We use the expansion
$$
({\bf {\hat a} \cdot {\hat b}})^p=p!\sum_{\ell=p,p-2,p-4,\ldots}\frac{2 \ell+1}{(p-\ell)!!(p+\ell+1)!!}~P_\ell({\bf {\hat a} \cdot {\hat b}})
$$ 
in the power series for the exponential
$$
e^{{\bf a \cdot b}}=\sum_{p=0}^\infty\frac{({\bf a \cdot b})^p}{p!}=\sum_{p=0}^\infty(ab)^p \frac{({\bf {\hat a} \cdot {\hat b}})^p}{p!}
$$
to get 
\begin{eqnarray}
e^{{\bf a \cdot b}}&=&\sum_{p=0}^\infty (ab)^p\sum_{\ell=p,p-2,p-4,\ldots}\frac{2 \ell+1}{(p-\ell)!!(p+\ell+1)!!}~P_\ell({\bf {\hat a} \cdot {\hat b}})\\ \nonumber
~&=&\sum_{\ell=0}^\infty(2\ell+1)(ab)^\ell P_\ell({\bf {\hat a} \cdot {\hat b}})~\sum_{n=0}^{\infty}\frac{(a^2 b^2)^n}{(2n)!!(2 n+2 \ell +1)!!}.
\end{eqnarray}
The Legendre polynomial can be decomposed  by means of the spherical harmonic addition theorem
$$
(ab)^\ell P_\ell({\hat a}\cdot{\hat b})\kk=\kk \frac{4\pi}{2\ell+1}(ab)^\ell \sum_{m=-\ell}^{\ell}Y^\ell_{-m}({\hat a})Y^\ell_{m}({\hat b})
\kk=\kk \frac{4\pi}{2\ell+1}\sum_{m=-\ell}^{\ell}(-1)^m{\cal Y}_{-m}^\ell({\bf a}){\cal Y}_{m}^\ell({\bf b}),
$$
leading to
\beq
e^{{\bf a\cdot b}}\kk=\kk 4\pi \sum_{n,\ell=0}^\infty\frac{[(a_x^2+a_y^2+a_z^2)(b_x^2+b_y^2+b_z^2)]^n}{(2n)!!(2n+2\ell+1)!!}
\sum_{m=-\ell}^{\ell}(-1)^m{\cal Y}_{-m}^\ell({\bf a}){\cal Y}_{m}^\ell({\bf b})
\eeq
which is Equation (2.14).

\section{The normalization factor for 3-dimensional harmonic oscillator states (Equation (2.15b))}

We start with 
\beq
{\cal Y}^\ell_\ell(c^+_x, c^+_y,c^+_z)|0>~=~\frac{(-1)^\ell}{2^{\ell/2}\ell!}\sqrt{\frac{(2\ell+1)!}{4\pi}}
(c^+_x+ic^+_y)^\ell|0>.
\eeq
Define
\beq
c^+_+~\equiv~\frac{c^+_x+ic^+_y}{\sqrt{2}}~~,~~~~~~~~~~c_+~\equiv~\frac{c_x-ic_y}{\sqrt{2}}~,
\eeq
for which
\beq
[c_+,c^+_+]~=~1~,~~~~~~~~~c_+|0>~=~0,
\eeq
so that $<0|(c_+)^\ell(c^+_+)^\ell|0>~=~\ell!$. Thus
\beq
<0|\left[{\cal Y}^\ell_\ell(c^+_x, c^+_y,c^+_z)\right]^+~{\cal Y}^\ell_\ell(c^+_x, c^+_y,c^+_z)|0>~=~
\left[\frac{(-1)^\ell}{2^{\ell/2}\ell!}\sqrt{\frac{(2\ell+1)!}{4\pi}}\right]^2\ell!~=~\frac{(2\ell+1)!}{4\pi\cdot 2^\ell \ell!}
\eeq

Now define the spherically symmetric operator $H^+\equiv (c^+_x)^2+(c^+_y)^2+(c^+_z)^2$. When it acts on a state it raises the number
of quanta by two, without changing the state's rotational transformation properties. Using (2.12), it can be shown that
\beq 
[H,H^+]~=~4N+6,
\eeq
where $N\equiv c^+_xc_x+c^+_yc_y+c^+_zc_z$ is the number operator. Then
\begin{eqnarray*}
&& H^n(H^+)^n{\cal Y}^\ell_\ell(c^+_x, c^+_y,c^+_z)|0>~=~H^{n-1}HH^+(H^+)^{n-1}{\cal Y}^\ell_\ell(c^+_x, c^+_y,c^+_z)|0>\\
&=&H^{n-1}H^+H(H^+)^{n-1}{\cal Y}^\ell_\ell(c^+_x,c^+_y,c^+_z)|0>+(4(2n-2+\ell)+6)H^{n-1}(H^+)^{n-1}
{\cal Y}^\ell_\ell(c^+_x, c^+_y,c^+_z)|0>
\end{eqnarray*}
If $H$ is repeatedly commuted past factors of $H^+$, the result is
\bea
&&H^n(H^+)^n{\cal Y}^\ell_\ell(c^+_x, c^+_y,c^+_z)|0>\nonumber\\
&=&\left[\{4(2n-2+\ell)+6\}+\{4(2n-4+\ell)+6\}+ \cdots \{4\ell+6\}\right]H^{n-1}(H^+)^{n-1}{\cal Y}^\ell_\ell(c^+_x,c^+_y,c^+_z)|0>\nonumber\\
&+&H^{n-1}(H^+)^{n}H{\cal Y}^\ell_\ell(c^+_x,c^+_y,c^+_z)|0>
\eea
But the last term in (B6) vanishes because $H{\cal Y}^\ell_\ell(c^+_x,c^+_y,c^+_z)|0>$ would have to be a state with $\ell-2$ quanta, and angular 
momentum quantum nunbers $(\ell,\ell)$, which is impossible. If we sum the arithmetic series in (B6), we get
\beq
H^n(H^+)^n{\cal Y}^\ell_\ell(c^+_x, c^+_y,c^+_z)|0>\kk=\kk 2n(2n+2\ell+1)H^{n-1}(H^+)^{n-1}{\cal Y}^\ell_\ell(c^+_x,c^+_y,c^+_z)|0>
\eeq
Iteration of (B7) gives
$$
H^n(H^+)^n{\cal Y}^\ell_\ell(c^+_x, c^+_y,c^+_z)|0>\kk=\kk \frac{(2n)!!(2n+2\ell+1)!!}{(2\ell+1)!!}{\cal Y}^\ell_\ell(c^+_x, c^+_y,c^+_z)|0>
$$
If this is combined with (B4), the result is
$$
<0|\left[(H^+)^n{\cal Y}^\ell_\ell(c^+_x, c^+_y,c^+_z)\right]^+~(H^+)^n{\cal Y}^\ell_\ell(c^+_x, c^+_y,c^+_z)|0>~=~
\frac{(2n)!!(2n+2\ell+1)!!}{4\pi},
$$
so that a normalized 3-dimensional harmonic oscillator eigenstate with angular momentum $\ell$, angular momentum $z$-component $\ell$, and 
$2n+\ell$ quanta can be written
\beq
(-1)^n\sqrt{\frac{4\pi}{(2n)!!(2n+2\ell+1)!!}}\left((c_x^+)^2+(c_y^+)^2+(c_z^+)^2\right)^n{\cal Y}^\ell_\ell(c^+_x, c^+_y,c^+_z)|0>
\eeq
A factor $(-1)^n$ has been inserted to yield phases for the eigenstates consistent with those used by Brody and Moshinsky (\cite{MB}).

Since the scalar products $<\Psi^\ell_m|\Psi^\ell_m>~(m=\ell,\ell-1\,\cdots -\ell)$ are independent of $m$, it follows that the normalization factor given in (B8) is also applicable if ${\cal Y}^\ell_\ell$ is replaced by ${\cal Y}^\ell_m$. This completes the justification of Eq.(2.15b).

\section{ Integration along the hyperbolic orbits}
\medskip

In order to calculate the Fourier transforms of the expressions for ${\bf F}(t)$
and ${\bf G}(t)$ given in eqs. (3.18), we need the following integrals
\beq
\int_{-\infty}^\infty e^{i\omega t}\frac{y'(t)}{[(r'(t)]^3}dt,~~~~~
\int_{-\infty}^\infty e^{i\omega t}\frac{z'(t)}{[(r'(t)]^3}dt,~~~~~
\int_{-\infty}^\infty e^{i\omega t}\frac{{\dot y'}(t)}{r'(t)}dt,~~~~~
\int_{-\infty}^\infty e^{i\omega t}\frac{{\dot z'}(t)}{r'(t)}dt
\eeq
It is convenient to express these integrals in terms of a parametric representation of the projectile orbit \cite{AW}:
\bes
\bea
y'(w)&=&a[\cosh(w)+\epsilon]\\
z'(w)&=&a\sqrt{\epsilon^2-1}\sinh(w)\\
t(w)&=&\frac{a}{v}[\epsilon \sinh(w)+w]
\eea
\ees
where $w$ is a parameter $(-\infty\leq w\leq \infty)$, $v$ is the asymptotic projectile speed, and $a$ and $\epsilon$ are defined in §III.2.
From (C2), it follows that
\begin{eqnarray*}
r'(w)&=&a(1+\epsilon\cosh(w))\\
dt&=&\frac{r'}{v}dw\\
{\dot y'}dt&=&dy'~=~a\sinh(w)dw\\
{\dot z'}&=&dz'~=~a\sqrt{\epsilon^2-1}\cosh(w)dw
\end{eqnarray*}
We can then see that
\bea
\int_{-\infty}^\infty e^{i\omega t}\frac{z'(t)}{[(r'(t)]^3}dt&=&\frac{\sqrt{\epsilon^2-1}}{av}\int_{-\infty}^\infty
e^{i\frac{wa}{v}[\epsilon \sinh(w)+w]}\frac{\sinh(w)}{(1+a\cosh(w))^2}dw\nonumber\\
&=&\frac{\sqrt{\epsilon^2-1}}{av}\int_{-\infty}^\infty e^{i\xi[\epsilon \sinh(w)+w]}\frac{d}{dw}\left(-\frac{1}{\epsilon}
(1+\epsilon\cosh(w))^{-1}\right)dw\nonumber\\
&=&\frac{\sqrt{\epsilon^2-1}}{\epsilon av}\int_{-\infty}^\infty (1+\epsilon\cosh(w))^{-1}\frac{d}{dw}
\left(e^{i\xi[\epsilon \sinh(w)+w]}\right)dw\nonumber\\
&=&i\xi\frac{\sqrt{\epsilon^2-1}}{\epsilon av}\int_{-\infty}^\infty e^{i\xi[\epsilon \sinh(w)+w]}dw
\eea
This last integral can be cast in a more convenient form by displacing the integration path from the real $w$ axis to the line 
$w+i\pi/2$ ($-\infty\leq w\leq\infty$). We can do this without changing the value of the integral because the integrand has no poles
between these two lines. Then (C3) becomes
\bea
\int_{-\infty}^\infty e^{i\omega t}\frac{z'(t)}{[(r'(t)]^3}dt&=&i\xi\frac{\sqrt{\epsilon^2-1}}{\epsilon av}
\int_{-\infty}^\infty e^{i\xi[\epsilon(\frac{e^{w+i\pi/2}-{e^{-w-i\pi/2}}}{2})+w+i\pi/2]}dw\nonumber\\
&=&i\xi\frac{\sqrt{\epsilon^2-1}}{\epsilon av}e^{-\xi\pi/2}\int_{-\infty}^\infty e^{-\xi\epsilon\cosh(w)+i\xi w}dw\nonumber\\
&=&2i\xi\frac{\sqrt{\epsilon^2-1}}{\epsilon av}e^{-\xi\pi/2}\int_0^\infty e^{-\xi\epsilon\cosh(w)}\cos(\xi w)dw
\eea

To proceed, we define $I(\epsilon)$ by
\begin{eqnarray*}
I(\epsilon)&=&\int_{-\infty}^\infty e^{i\xi[\epsilon \sinh(w)+w]}\frac{\sinh(w)}{(1+a\cosh(w))^2}dw\\
&=&2i\frac{\xi}{\epsilon}e^{-\xi\pi/2}\int_0^\infty e^{-\xi\epsilon\cosh(w)}\cos(\xi w)dw
\end{eqnarray*}
Then a partial integration and some re-arrangement show that
\begin{eqnarray*}
\frac{d}{d\epsilon}\left[i\frac{\epsilon}{\xi}I(\epsilon)\right]&=&\int_{-\infty}^\infty e^{-\xi\epsilon\cosh(w)+w}
\frac{\epsilon+\cosh(w)}{(1+\epsilon\cosh(w))^2}dw\\
&=&\frac{d}{d\epsilon}\left[-2e^{-\xi\pi/2}\int_0^\infty e^{-\xi\epsilon\cosh(w)}\cos(w)dw\right]\\
&=&2\xi e^{-\xi\pi/2}\int_0^\infty e^{-\xi\epsilon\cosh(w)}\cos(w)\cosh(w)dw
\end{eqnarray*}
Thus
\bea
\int_{-\infty}^\infty e^{i\omega t}\frac{y'(t)}{[(r'(t)]^3}dt&=&\frac{1}{av}\int_{-\infty}^\infty e^{i\xi[\epsilon\sinh(w)+w]}
\frac{\epsilon+\cosh(w)}{(1+\epsilon\cosh(w))^2}dw\nonumber\\
&=&\frac{2\xi}{av}e^{-\xi\pi/2}\int_0^\infty e^{-\xi\epsilon\cosh(w)}\cos(w)\cosh(w)dw
\eea
This same shift of the integration path leads to
\bea
\int_{-\infty}^\infty e^{i\omega t}\frac{{\dot y'}(t)}{r'(t)}dt&=&\int_{-\infty}^\infty e^{i\xi[\epsilon\sinh(w)+w]}
\frac{\sinh(w)}{1+\epsilon\cosh(w)}dw\nonumber\\
&=& 2ie^{-\xi\pi/2}\int_0^\infty e^{-\xi\epsilon\cosh(w)}\cosh(w)\frac{\cos(\xi w)+\epsilon\sinh(w)\sinh(\xi w)}{1+(\epsilon\sinh(w))^2}dw
\eea
and
\bea
\int_{-\infty}^\infty e^{i\omega t}\frac{{\dot z'}(t)}{r'(t)}dt&=&\sqrt{\epsilon^2-1}\int_{-\infty}^\infty e^{i\xi[\epsilon\sinh(w)+w]}
\frac{\cosh(w)}{1+\epsilon\cosh(w)}dw\nonumber\\
&=&2\sqrt{\epsilon^2-1}e^{-\xi\pi/2}\int_0^\infty e^{-\xi\epsilon\cosh(w)}\sinh(w)\frac{-\sin(\xi w)+\epsilon\sinh(w)\cosh(\xi w)}{1+(\epsilon\sinh(w))^2}dw
\eea

If the four integrals (C4, C5, C6, C7) are used in (3.18) and (3.8b), the result is the set of expressions (3.19) for $\alpha_y$
and $\alpha_z$.

\clearpage
\normalsize
\begin{figure}
\includegraphics{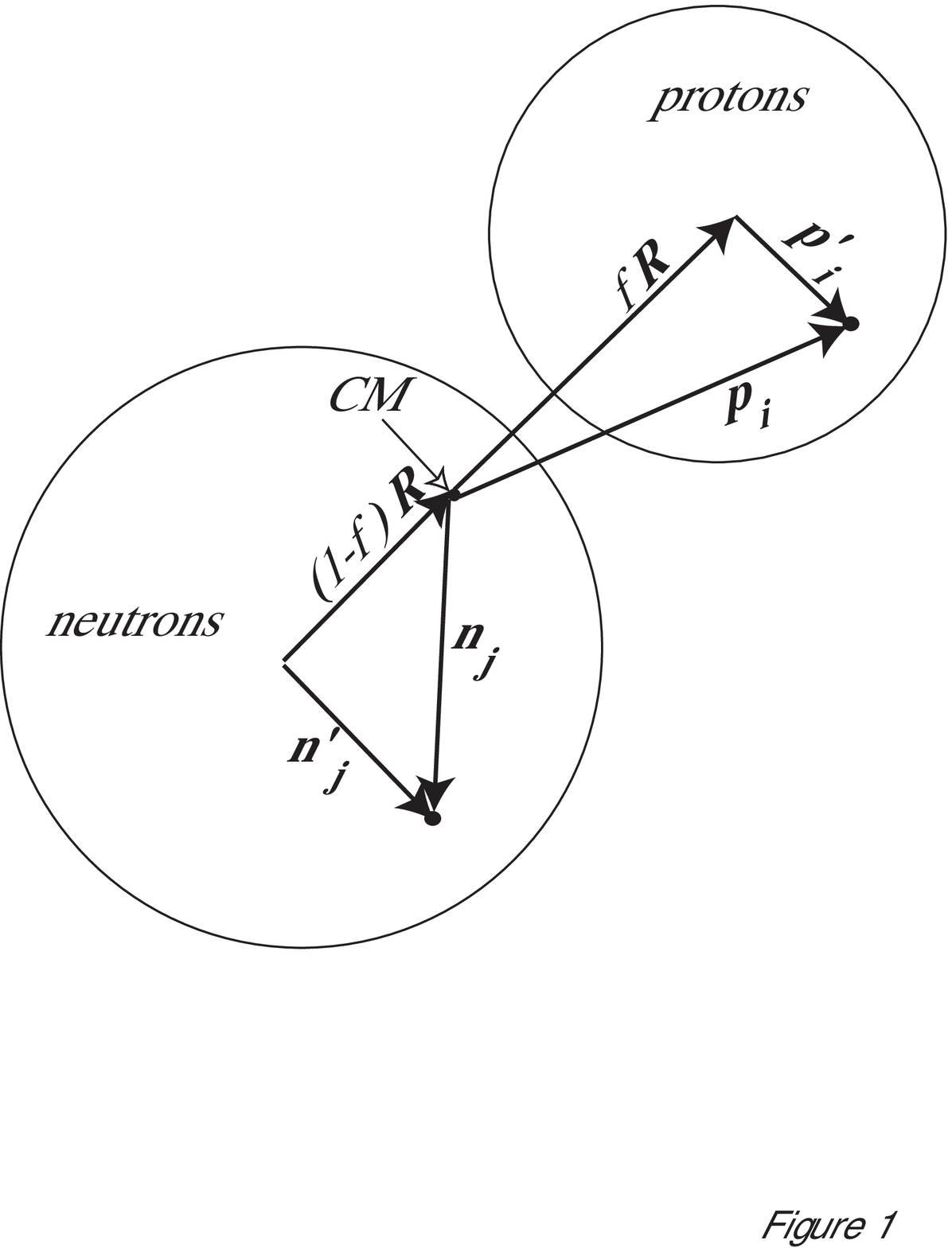}
\caption{The vector ${\bf R}$ connects the neutron mass center to the proton mass center. ${\bf p}_i$ locates the i$^{th}$ proton relative to the nuclear mass center, whereas ${\bf p'}_i$ locates it relative to the proton mass center. $f$ is the ratio $N_{\rm T}/A_{\rm T}$.}
\end{figure}
\begin{figure}
\includegraphics{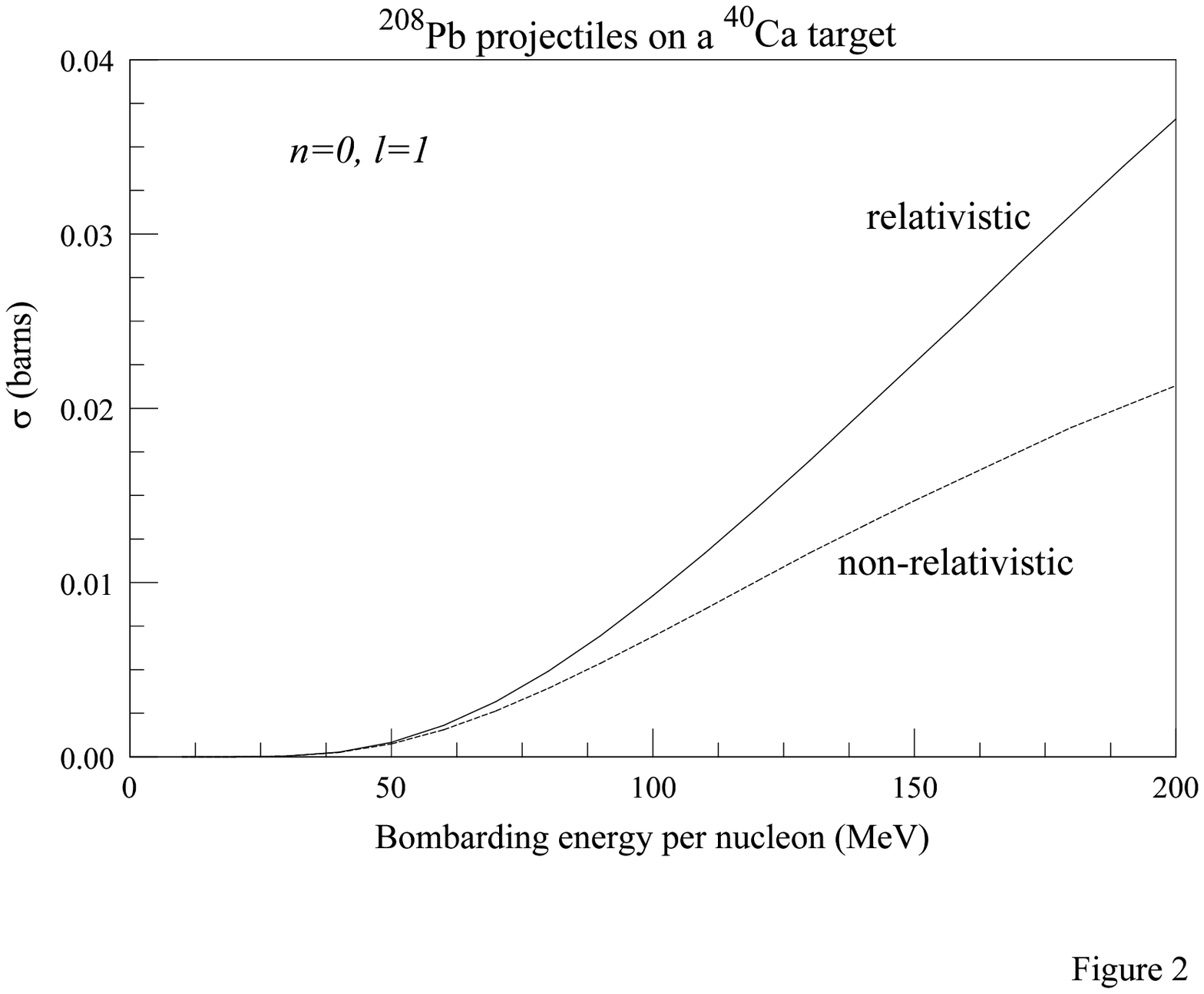}
\caption{Comparison of results of relativistic and non-relativistic calculations of the cross-section for excitation of the one-phonon GDR level in $^{40}$Ca, due to Coulomb excitation by $^{208}$Pb projectiles.}
\end{figure}
\begin{figure}
\includegraphics[scale=.7]{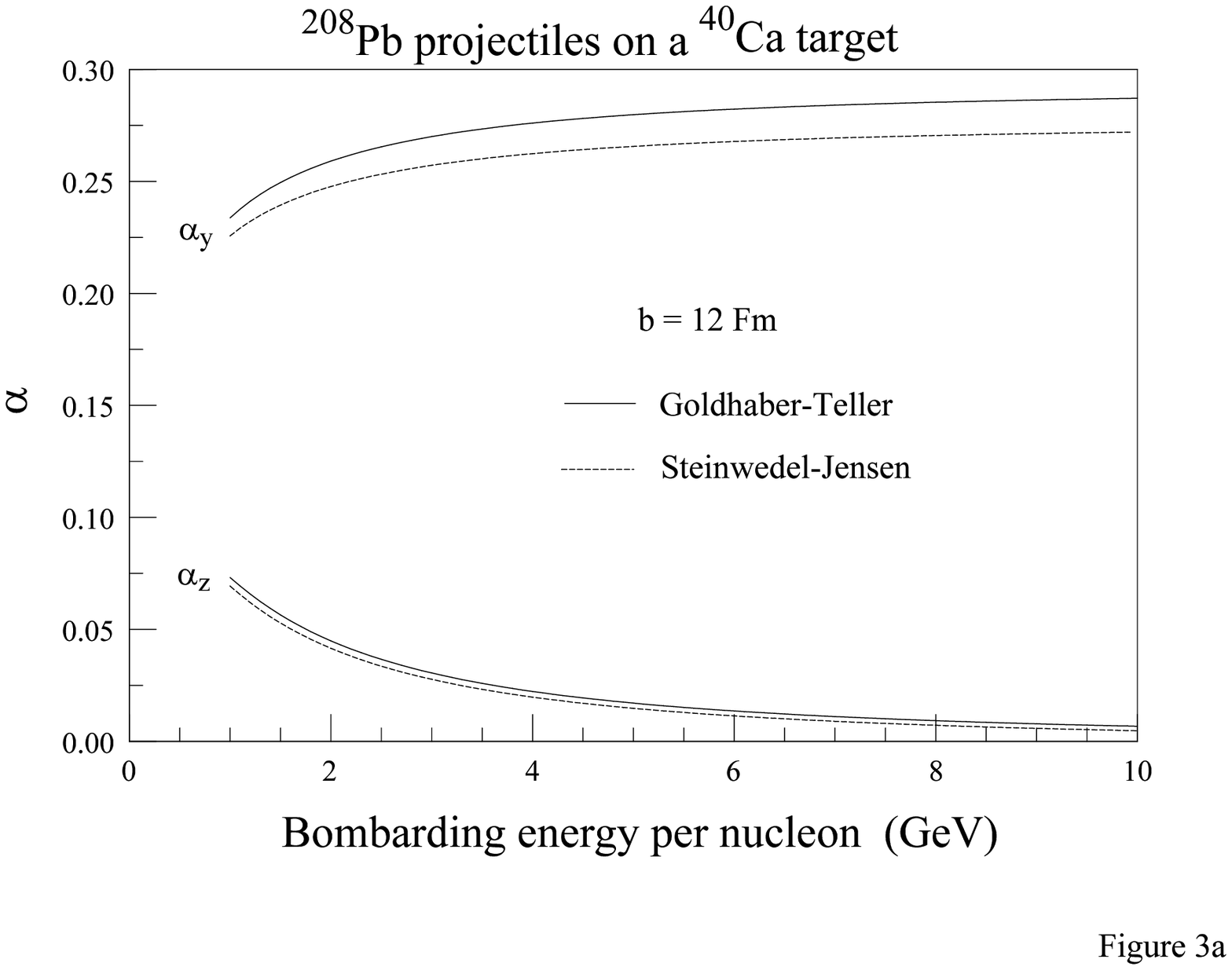}\hfill\includegraphics[scale=.7]{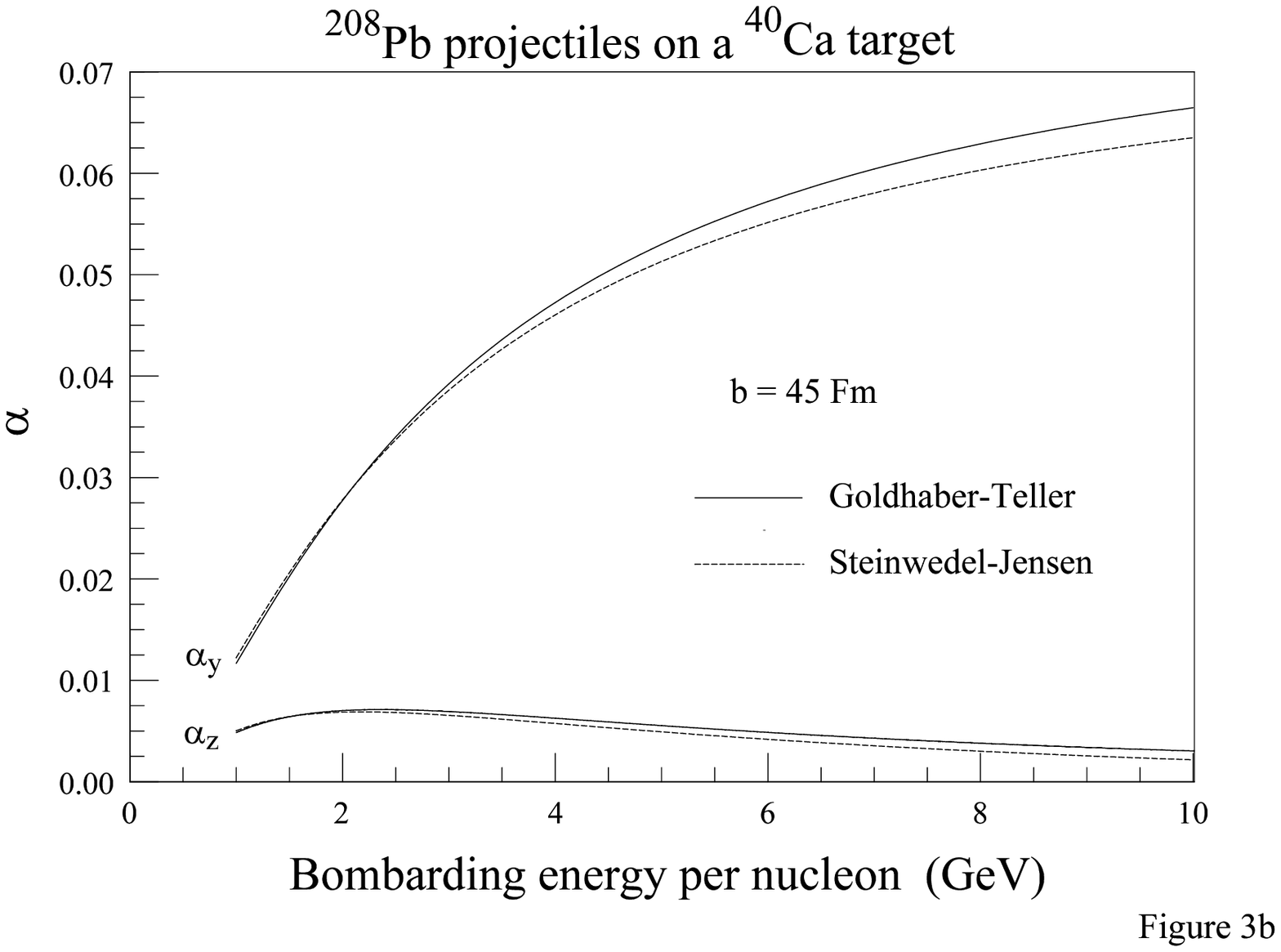}
\caption{Comparison of calculations of $(\alpha_y,\alpha_z)$ using the Goldhaber-Teller and Steinwedel-Jensen models of the GDR. a) $b=12$ fm; b) $b=45$ fm.}
\end{figure}
\begin{figure}
\includegraphics{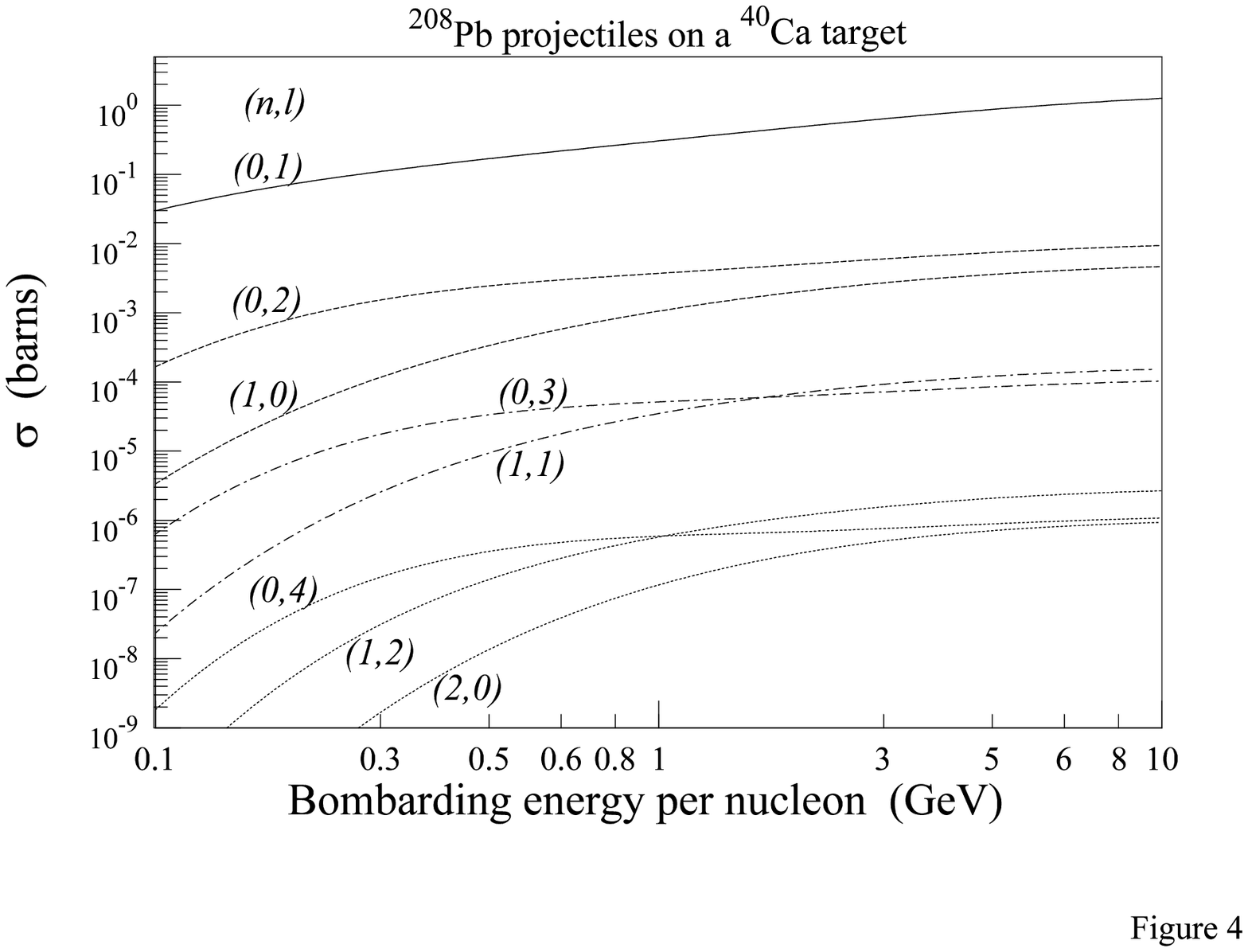}
\caption{Excitation cross-sections for various $(n,\ell)$ levels of the GDR in $^{40}$Ca, due to Coulomb excitation by $^{208}$Pb projectiles. Levels with the same number of phonons are indicated by the same type of line. These calculations are done using the Goldhaber-Teller description of the GDR.}
\end{figure}


\begin{thebibliography}{99}

\bibitem{GT} M. Goldhaber and E. Teller, Phys. Rev. {\bf 74}, 1046 (1948).

\bibitem{SJ} H. Steinwedel and J.H.D. Jensen, Phys. Rev. {\bf 79}, 1019 (1950).

\bibitem{LA} E.G. Lanza,  M.V. Andres, F. Catara, Ph. Chomaz and C. Volpe, Nucl.Phys. {\bf A613}, 445 (1997).

\bibitem{CA} E.G. Lanza, F. Catara, M.V. Andres, and Ph. Chomaz, M. Fallot, and J.A. Scarpaci, Phys. Rev. C {\bf 79} 054615 (2009).

\bibitem{BP} C.A. Bertulani and V.Yu. Ponomarev, Phys. Rep. {\bf 331}, 139 (1999).

\bibitem{ABE} T. Aumann, P.F. Bortignon and H. Emling, Annu. Rev. Nucl. Part. Sci. {48} 351 (1998).

\bibitem{CF} Ph. Chomaz and N. Frascaria, Phys. Rep. {\bf 252}, 275 (1995).

\bibitem{BE} C.A. Bertulani and G.Baur, Physics Reports {\bf 163} 299 (1988). 

\bibitem{ME} E. Merzbacher, {\it Quantum Mechanics}, Second Edition, J. Wiley \& Sons, New York (1970).

\bibitem{AW} K. Alder and A. Winther, {\it Electromagnetic Excitation}, North Holland, Amsterdam (1975).

\bibitem{WA} A. Winther and K. Alder, Nuclear Physics {\bf A319} (1979) 518.

\bibitem{JA} J.D. Jackson, {\it Classical Electrodynamics}, Third Edition, J. Wiley \& Sons, New York (1999).

\bibitem{BZ} B.F. Bayman and F. Zardi, Phys. Rev. C {\bf 59} 2189 (1999).

\bibitem{ZB} B.F. Bayman and F. Zardi, Phys. Rev. C {\bf 74}, 024905 (2006).

\bibitem{GR} W. Greiner and and J.A. Maruhn, {\it Nuclear Models}, Springer-Verlag, Berlin (1996).

\bibitem{BM} A. Bohr and B.R. Mottelson, {\it Nuclear Structure}, Vol II, p.476, W.A. Benjamin, Reading (1975). 

\bibitem{MB} T.A. Brody and M. Moshinsky, {\it Tables of Transformation Brackets}, Monographias del Instituto de Fisica, Mexico, (1960).



\end{thebibliography}
\end{document}